\documentclass[%
 reprint,
preprintnumbers,
nofootinbib,
amsmath,amssymb,aps,superscriptaddress
]{revtex4-2}
\counterwithout{equation}{section}
\usepackage{amsmath,float,graphicx,placeins,amssymb,multirow,pgfplots,pgfplotstable}
\usepackage{empheq}
\usepackage{booktabs}
\usepackage{tikz}\usepackage{amsmath}
\usepackage[most]{tcolorbox} 
\usepackage{amsmath,amssymb,amsfonts,bm}
\usepackage{mathtools}
\usepackage{physics}
\usepackage{microtype}
\usepackage{xcolor}
\usepackage{comment}
\usepackage{enumerate,slashed,cancel,comment,color,bbm,graphicx,rotating,placeins,mathtools,multirow,hhline}
\usepackage[normalem]{ulem}
\usepackage{array}
\usepackage{hyperref}
\usepackage{color}
\usepackage{booktabs}
 \hypersetup
 {
   colorlinks,
   linkcolor={blue!80!black},
   citecolor={blue!70!black},
   urlcolor={blue!70!black}
 }

\usepackage{hyperref}
\usepackage{tabularx}

\newcommand{\be}{\begin{equation}}
\newcommand{\ee}{\end{equation}}
\newcommand{\bea}{\begin{eqnarray}}
\newcommand{\eea}{\end{eqnarray}}
\newcommand{\ba}{\begin{aligned}}
\newcommand{\ea}{\end{aligned}}

\newcommand{\epsilonH}{\varepsilon}
\newcommand{\nn}{\nonumber\\}

\pgfplotsset{compat=1.18} 
\begin{document}
\title{Wilsonian Cosmology: de Sitter (in)Stability}
\begin{abstract} 
We develop a (Wilsonian) functional-renormalisation-group framework for scalar cosmology in which quantum fluctuations of a scalar field are coarse-grained on cosmological spacelike hypersurfaces. Integrating out quantum fluctuations with wavelengths smaller than the Hubble radius $H(t)^{-1}$, we obtain an effective scalar potential $U(\phi(t),H(t))$ that evolves in time. We derive a non-perturbative flow equation for this potential, together with the coupled set of (modified) Friedmann equations. 
We then apply this formalism to the simplest possible case of a de Sitter vacuum when the scalar field is at rest, in two 
situations which satisfy exactly our flow equation: $(i)$ A flat potential, for which we find that the only pure de Sitter solution is unstable and corresponds to a saddle point. $(ii)$ A quadratic potential with curvature $m^2>0$, for which we find that the presence of the mass term stabilises the system. The latter case leads to a de Sitter attractor either at the Hubble scale when the mass is larger than the Hubble scale at initial time, or by introducing a new attractor located at $H=m$ in the case the mass is smaller, which may be particularly relevant to the phenomenological study of dark energy and inflation theories. 
\end{abstract}

\author{Jean Alexandre}
\email[Email address: ]{jean.alexandre@kcl.ac.uk}

\author{Lucien Heurtier}
\email[Email address: ]{lucien.heurtier@kcl.ac.uk}

\affiliation{Theoretical Particle Physics and Cosmology, King’s College London,\\ Strand, London WC2R 2LS, United Kingdom}

\author{Silvia Pla}
\email[Email address: ]{silvia.pla-garcia@tum.de}

\affiliation{Physik-Department, Technische Universit\"at M\"unchen, James-Franck-Str., 85748 Garching, Germany}

\maketitle

\tableofcontents

\section{Introduction}

Particle physics in cosmology is commonly described within a
semiclassical framework, in which quantum fields propagate on a
classically evolving spacetime. This approximation is particularly
transparent during the hot Big Bang era, where microscopic and cosmological scales are widely separated. For relativistic particles in
a thermal plasma with temperature $T$, the typical microscopic wavelength is
\(\lambda_{\rm th}\sim T^{-1}\), whereas during radiation domination the
Hubble radius is of order
\(\lambda_H=H^{-1}\sim M_p/T^2 \gg\lambda_{\rm th}\): particle interactions
take place on scales much smaller than the Hubble radius and are therefore locally insensitive to spacetime
curvature. The expanding geometry nevertheless acts continuously on the
plasma through the dilatation of physical distances, redshift of momenta, dilution of particle number densities, and subsequent cooling of the thermal bath as
\(T\propto a^{-1}\), where $a$ denotes the scale factor. As a result, even though the underlying quantum field theory is locally insensitive to
the curvature of the Universe in this regime, its observable predictions
are thus definitely sensitive to cosmic expansion through the evolving temperature of the plasma. This temperature dependence enters the microscopic description in two ways. Quantum effects
generate thermal contributions to the
effective potential, and the characteristic energy scale
of microscopic reactions is set by the temperature. Since renormalisation-group-improved calculations are naturally organised around the characteristic physical scale of the process, they naturally involve masses and
couplings evaluated at a scale \(\mu\sim T\). As the Universe expands and
cools, particle-physics observables therefore substantially evolve through both the ordinary running of the couplings evaluated at the thermal scale and genuine finite-temperature effects. Cross sections, decay rates, effective masses and the shape of the scalar potential can all change substantially over the cosmological history, even though the spacetime itself is treated classically
\cite{Quiros:1999jp,Laine:2016hma}.

The thermal Universe thus provides a familiar example of a more general principle: an evolving physical scale in the cosmological environment can induce an unavoidable evolution of the effective quantum theory used to describe matter. The situation becomes conceptually less straightforward outside thermal equilibrium. 
At zero temperature there is no thermal scale \(T\) with which to organise the effective theory,
and the energy, momentum and occupation number of a field need not be directly related to the Hubble rate. A mode with physical momentum \(k/a\gg H\) evolves on a timescale much shorter than the expansion time and can ordinarily be treated as a local quantum excitation. By contrast, modes with frequencies comparable to or smaller than \(H\) are sensitive to the time dependence of the background and may require a qualitatively different long-wavelength description.

{This separation plays an essential role in several apparently distinct cosmological phenomena. It underlies the coherent-wave description of ultralight dark matter, for which large occupation numbers allow a classical field description on macroscopic scales~\cite{Marsh:2015xka,Hui:2016ltb}; it plays a central role in gravitational particle production~\cite{Parker:1968mv,Ford:2021syk}; and it lies at the heart of stochastic inflation, where modes are continuously transferred from a short-wavelength quantum sector to a coarse-grained long-wavelength sector as they approach the Hubble scale~\cite{Starobinsky:1986fx,Starobinsky:1994bd}. These examples illustrate the special role played by the Hubble scale in organising cosmological fluctuations: $H$ provides a distinguished, time-dependent dynamical scale separating short-wavelength modes whose evolution is approximately local from long-wavelength modes whose dynamics is sensitive to the cosmological expansion.}

In most cosmological applications, one specifies a renormalised action at
a chosen reference scale and subsequently evolves the fields and their
quantum fluctuations on the expanding background. Cosmic expansion
redshifts the physical momenta of the modes and may carry them from the
sub-Hubble to the super-Hubble regime, but the masses, couplings and
potential entering the action are not themselves evolved with the Hubble
scale.

From a Wilsonian perspective, however, the long-wavelength sector should
be described by an effective theory obtained after integrating out
short-wavelength fluctuations above a physical cutoff \(\kappa\). The
parameters of this effective theory necessarily depend on the position
of that cutoff, since changing \(\kappa\) changes which fluctuations have
already been incorporated into its couplings and potential. In a
cosmological setting, the Hubble rate provides a distinguished physical
scale separating modes that evolve locally on sub-Hubble scales from the
long-wavelength modes retained in the effective description. Identifying
\(\kappa\) with \(H\), the set of integrated-out fluctuations therefore
changes as the Universe evolves. The Hubble scale does not merely
determine which modes belong to the long-wavelength sector; it also
determines the effective theory that governs their interactions.

This observation motivates the framework developed in this work. 
Following the approach we developed in Ref.~\cite{Alexandre:2025ixz}, 
and inspired by the effective-average-action approach of Wetterich
\cite{Wetterich:1992yh,Wetterich:1989xg,Wetterich:2001kra}, we
construct a spatially coarse-grained effective action at each
cosmological time and identify its physical Wilsonian scale with the
Hubble rate,
\[
\mbox{Wilsonian running scale}~=H(t).
\]
As we will see, the mismatch between the spacetime dimension and the dimension in which modes are coarse-grained introduces both a spatial regulator and a temporal frequency cutoff, fixed by requiring that the adiabatic flat-spacetime limit reproduces the
standard four-dimensional functional-renormalisation-group equation.
The resulting effective scalar potential is a function
\(U(\phi,H)\), whose evolution must be followed simultaneously with that
of the scalar background and the spacetime geometry. The dependence of
the effective theory on \(H\) modifies the energy density and pressure
entering the Friedmann equations, while the evolution of the Friedmann
background in turn determines the Wilsonian scale at which the theory is
defined. The effective scalar theory and the cosmological geometry
therefore form a single, self-consistent dynamical system.

The framework developed here is particularly suited to studying the fate of the de Sitter spacetime under quantum coarse-graining, and this is what we will focus on throughout this paper. De Sitter space plays a central role in cosmology, both as the idealised limit of an inflationary phase and as the simplest description of accelerated expansion driven by vacuum energy. Understanding its behaviour under quantum corrections is therefore an important question
, that we address here by asking whether constant-Hubble solutions of our equations act as attractive or repulsive fixed points for Wilsonian flows.

Infrared effects in de Sitter space have been studied using perturbative quantum field theory, stochastic inflation, semiclassical gravity and functional-renormalisation-group methods.  Perturbative calculations for light or massless scalar fields can contain contributions that grow with time and have sometimes been interpreted as indicating an instability of de Sitter space~\cite{Ford:1984hs}. This conclusion is not universal, however. For self-interacting light scalars on a prescribed de Sitter background, stochastic and nonperturbative approaches show that the large late-time contributions can be resummed, leading in many cases to equilibration and to the generation of infrared scales \cite{Starobinsky:1994bd,Gorbenko:2019rza,Guilleux:2015pma,Cespedes:2023aal}. Once the geometry is allowed to respond, the picture becomes more model dependent: semiclassical studies find outcomes that depend on the matter content, the quantum state and the coarse-graining prescription \cite{Perez-Nadal:2007yxe,Markkanen:2016aes,Moreau:2018lmz,
Moreau:2018ena,Markkanen:2017abw}.

Stochastic inflation and Wilsonian approaches both rely on a dynamical separation between short- and long-wavelength fluctuations. In an expanding spacetime, physical momenta are continuously redshifted and modes cross the coarse-graining scale, but the two frameworks encode the resulting infrared dynamics differently. In stochastic inflation, this mode transfer leads to a probabilistic description of the coarse-grained long-wavelength field~\cite{Starobinsky:1986fx}. Functional renormalisation-group approaches instead track how the effective action, and in particular the effective potential, changes as fluctuations are progressively integrated out~\cite{Serreau:2013eoa,Guilleux:2015pma,Guilleux:2016oqv,Kaya:2013bga,Banerjee:2022xvi}. The relation between these descriptions has been explored from different perspectives~\cite{Prokopec:2017vxx,Cespedes:2026fdp}. In the present work we adopt the Wilsonian perspective, but identify the coarse-graining scale with the dynamical Hubble rate itself. As a result, the Hubble-dependent effective potential and the homogeneous spacetime geometry must be evolved self-consistently, which is the central distinction of our framework.

The paper is organised as follows. In Sec.~II we derive the spatially
coarse-grained functional flow for a scalar potential in the presence of
a time-dependent background for: the field, the FLRW metric and the running scale. 
We present the main steps only, focusing on the approach and the assumptions, 
while the details are covered in Appendix A.
In Sec.~III we identify the physical running scale with the Hubble rate and we derive the corresponding
modified Friedmann equations, which consistently satisfy the continuity equation. 
In Sec.~IV we study the resulting Wilsonian dynamics for field-independent and quadratic scalar
potentials, determine their constant-Hubble equilibria and analyse their
local and global stability. 
The concluding Sec. V is left for a discussion on the potential relevance 
of our results to the description of a transient phantom-like behaviour, 
arising from the non-monotonic evolution of the Hubble rate.
Finally, Appendix A provides the details for the derivation of our flow equation, 
and Appendix B reviews the usual
adiabatic expansion based on the Schwinger proper-time approach. This allows
us to compare our Wilsonian potential with the one-loop effective
potential and to clarify their similarities and differences.

\section{Exact Renormalisation Group Approach}

We consider the FLRW metric 
\be
ds^2=-dt^2+a^2(t)(d\vec x)^2~,
\ee
and for each time slice, we perform a coarse graining in space, where ultraviolet (UV) and infrared (IR) modes of the inflaton field are separated by a time-dependent comoving scale $k(t)$, which will eventually be related to the Hubble rate. We sketch here the steps of the derivation, for which details can be found in Appendix (\ref{derivation})

\subsection{Space-average effective action}

The inflaton field is decomposed as
\be
\Phi=\phi(t)+\varphi(t,\vec x)~,
\ee
where $\phi$ is a classical field and $\varphi$ is a quantum field, whose three-dimensional Fourier transform is defined with the comoving momentum $\vec p$
\be
\varphi(t,\vec p)=\int d^3x~\varphi(t,\vec x)e^{i\vec p\cdot\vec x}~.
\ee
In what follows we denote
\be
\int_p\equiv\int\frac{d^3p}{(2\pi)^3}~.
\ee
Following Wetterich approach to Wilsonian renormalisation, but with a coarse graining in space,  
we integrate over Fourier modes $\varphi(t,\vec p)$ weighted by a cutoff function $C_k(p^2)$ in the path integral, with $p=|\vec p|$. 
The scalar field bare action is then
\bea\label{eq:action}
&&S_k[\Phi]\\
&=&\int dt\sqrt{-g}\int d^3x\left(-\frac{g^{\mu\nu}}{2}\partial_\mu\Phi\partial_\nu\Phi-\frac{\xi}{2}R\Phi^2-U_i(\Phi)\right)\nn
&&-\frac{1}{2}\int dt \sqrt{-g}\int_p\varphi(t,\vec p)\varphi(t,-\vec p)~\Big(C_k(p^2)-i\varepsilon\Big)~,\nonumber
\eea
where the curvature scalar is
\be
R=6(2H^2+\dot H)~,
\ee
and $U_i$ is the bare potential, corresponding to the initial condition for the Wilsonian flow towards the IR.
The "$i\varepsilon$" prescription ($\varepsilon>0$) ensures that the path integral converges (see below).
The cutoff function plays the role of a mass of the order $k^2$ for IR modes (with $p^2<k^2$), 
and is suppressed for UV modes (with $p^2>k^2$). 
When defining the one-particle-irreducible (1PI) effective action $\Gamma_k$, UV modes are then integrated out without any constraint, 
while IR modes are mainly frozen, leading to the {\em average effective action}~\cite{Wetterich:1989xg}. By construction, one imposes $\lim_{k\to0} C_k(p^2)=0$ which implies that, 
in the deep infrared limit $k\to0$, $\Gamma_k$ becomes the usual 1PI effective action.

The partition function for the quantum field $\varphi$ on the background $\phi(t)$ is 
\bea \label{eq:eq00}
&&Z_k[j]\equiv \exp\Big(iW_k[j]/\hbar\Big)\\
&=&\int{\cal D}[\varphi]\exp\left((i/\hbar)S_k[\Phi]+(i/\hbar)\int dt \sqrt{-g}\int d^3x ~j\varphi\right)~.\nonumber
\eea
The background fluctuation field with momentum $\vec p$ is
\be
\varphi_b(t,\vec p)=\left<\varphi(t,\vec p)\right>\equiv-\frac{i\hbar}{\sqrt{-g}Z_k}\frac{\delta Z_k}{\delta j(t,-\vec p)}~.
\ee
The 1PI effective action is defined as the Legendre transform of $W[j]$, with the cutoff function term removed,
\bea
&&\Gamma_k[\phi+\varphi_b]=W_k[j]-\int dt' \sqrt{-g}\int d^3x ~j\varphi_b \\
&&+\frac{1}{2}\int dt' \sqrt{-g}\int_p \varphi_b(t',\vec p)\varphi_b(t',-\vec p)(C_k-i\varepsilon)~,\nonumber
\eea
and $j$ should be understood as a functional of $\varphi_b$. 
The evolution of the effective action with $k(t)$ is obtained from the functional derivative
\bea\label{dGammadk}
\frac{\delta\Gamma_k}{\delta k(t)}&=&i\hbar\frac{\sqrt{-g}}{2}\int_p\int_q\delta(\vec p+\vec q)\int dt' \delta(t-t')\\
&\times&\Bigg(\sqrt{-g(t)}(C_k-i\varepsilon)\delta(\vec p+\vec q)\delta(t-t')\nn
&&~~~~~~~~~~~~~~~~~~-\frac{\delta^2\Gamma_k}{\delta\varphi_b(t,\vec p)\delta\varphi_b(t',\vec q)}\Bigg)^{-1}\partial_k C_k~,\nonumber
\eea
which is detailed in Appendix A.

\subsection{Local Potential Approximation}

At this stage we consider the Local Potential Approximation (LPA) where, for every scale $k$, we assume that the effective action has the same form as the bare action (without the cutoff function term), 
but with the initial potential $U_i(\Phi)$ replaced by the running potential $U_k(\Phi)$
\bea
&&\Gamma_k[\Phi]\\
&=&\int dt\sqrt{-g}\int d^3x\left(-\frac{g^{\mu\nu}}{2}\partial_\mu\Phi\partial_\nu\Phi-\frac{\xi}{2}R\Phi^2-U_k(\Phi)\right)~.\nonumber
\eea
For a vanishing source $j=0=\varphi_b$, the evolution of the effective action evaluated at the background field is then given by
\be\label{dGammadkbis}
\frac{1}{\sqrt{-g}}\frac{\delta\Gamma_k[\phi]}{\delta k(t)}=-V\partial_kU_k(\phi)~,
\ee
where $V$ is the space volume. Taking into account these different expressions and performing a Wick rotation, we finally obtain the evolution equation for the effective action 
\be\label{dkU}
\partial_k U_k(\phi)=\hbar\frac{T^{-1}}{2}\int_p{\cal D}_E^{-1}\Big(a^{-3}\partial_kC_k\Big)~,
\ee
where 
\be\label{DE}
{\cal D}_E=-\frac{d^2}{dt^2}-3H\frac{d}{dt}+\kappa^2-\xi R+ \partial^2_\phi U_\kappa(\phi)~,
\ee
and $\kappa$ is defined here with imaginary time. 
The cutoff frequency $T^{-1}$ is unavoidable, and is a consequence of the mismatch of dimensions: 
three-dimensional coarse-graining in four-dimensional spacetime.

We then choose the Litim cutoff \cite{Litim:2001up} for the physical scales $p/a$ and $\kappa\equiv k/a$
\be
C_k(p^2)=a^{-2}(k^2-p^2)\Theta(k^2-p^2)~,
\ee
in which case the evolution equation (\ref{dkU}) becomes 
\be\label{flowE}
\partial_\kappa U_\kappa(\phi)=\hbar\frac{aT^{-1}}{6\pi^2}{\cal D}_E^{-1}\left(a^{-1}\kappa^4\right)~.
\ee
At this point we note that we could use an adiabatic expansion in order to determine how ${\cal D}_E^{-1}$ acts on the right-hand side of Eq.(\ref{flowE}), 
but this would involve a series in time derivatives of $R$ and $\partial^2_\phi U_\kappa(\phi)$. 
Instead of expanding ${\cal D}_{E}^{-1}$ adiabatically, we act with $\mathcal{D}_{E}$ on both sides of \eqref{flowE}. After analytical continuation back to Lorentzian time, this yields
\be\label{flow0}
{\cal D}\left(T\partial_\kappa U_\kappa(\phi)\right)=\frac{\hbar\kappa^4}{6\pi^2}~,
\ee
where we define
\bea\label{defD}
{\cal D}&\equiv&\frac{d^2}{dt^2}+H\frac{d}{dt}+\kappa^2+{\cal M}^2\\
{\cal M}^2&\equiv&\partial^2_\phi U_\kappa(\phi)+(\xi-1/6)R~,\nonumber
\eea
and $\kappa$ is here defined with real time. 

We note that the running of the potential is purely of quantum origin: $\partial_\kappa U_\kappa(\phi)={\cal O}(\hbar)$.
We also stress that Eq.(\ref{flow0}) is non-perturbative, in the sense that it results from a partial resummation of all quantum corrections which occur 
within the LPA framework. This corresponds to all the Feynman graphs involving non-derivative interactions, which can be obtained from the interactions appearing in the initial potential, defined at some scale $\kappa=\Lambda$.

\subsection{Evaluation of the frequency cutoff}

The flow equation (\ref{flow0}) depends on the frequency cut off $T^{-1}$ which is a free parameter in the present derivation. 
One can determine its value though, by taking the flat spacetime limit, where $H=0$ and $\dot\kappa=\dot\phi=\dot T=0$. 
In this case the flow equation reads
\be
\partial_\kappa U_\kappa(\phi)=\frac{T^{-1}}{6\pi^2}\frac{\hbar\kappa^4}{\kappa^2+\partial^2_\phi U_\kappa(\phi)}~.
\ee
This corresponds to the usual four-dimensional flow equation if one identifies $T^{-1}$ with $3\kappa/16$. 
In what follows we will therefore assume this value for $T^{-1}$, and the flow equation is finally
\be\label{flow}
\boxed{{\cal D}\left(\kappa^{-1}\partial_\kappa U_\kappa(\phi)\right)=\frac{\hbar\kappa^4}{32\pi^2}}~,
\ee
where ${\cal D}$ is given in Eq.(\ref{defD}).
Note that a time-dependent cutoff is allowed in the context of this approach, since the coarse-graining is in space only.

According to the intuitive argument given in \cite{Senatore:2009cf}, 
quantum corrections to the inflaton propagator should not depend on the ratio of a comoving scale and a physical scale, 
otherwise they would involve terms containing the scale factor $a$ on its own, and increase in time in a de Sitter regime. 
Consistently with this argument, the flow equation (\ref{flow}) indeed depends only on physical scales, which are $\kappa, H, R$ and $\partial^2_\phi U$, and the resulting quantum corrections cannot involve the scale factor explicitly.

\subsection{Recovering the 1PI effective potential at one loop} \label{subsec:recovering1PI}

We show here that the solution of the flow equation (\ref{flow}) naturally recovers known results in the adiabatic limit, when $\dot H=\dot\kappa=\dot\phi=0$,
and therefore
\be
\partial_\kappa U_\kappa(\phi)=\frac{\hbar}{32\pi^2}\frac{\kappa^5}{\kappa^2+{\cal M}^2}~.
\ee
Since the running of the potential is of quantum origin, it can be written $U_\kappa(\phi)=U_i(\phi)+{\cal O}(\hbar)$, 
where $U_i(\phi)$ is the bare (non-running) potential, which is identified as the initial potential at some initial scale $\Lambda$. 
An expansion in powers of $\hbar$ leads then to 
\be
\partial_\kappa U_\kappa(\phi)=\frac{\hbar}{32\pi^2}\frac{\kappa^5}{\kappa^2+{\cal M}^2_i}+{\cal O}(\hbar^2)~,
\ee
where 
\be
{\cal M}^2_i=\partial^2_\phi U_i(\phi)+(\xi-1/6)R~.
\ee
The integration over $\kappa$ leads then to the known one-loop 1PI effective potential $U^{(1)}\equiv U_{\kappa=0}$ in terms of the bare potential 
$U_i\equiv U_{\kappa=\Lambda}$:
\bea
U^{(1)}(\phi)&=&U_i(\phi)-\frac{\hbar}{32\pi^2}\int_0^\Lambda \frac{\kappa^5~d\kappa}{\kappa^2+{\cal M}^2_i}\\
&=&U_i(\phi)+\frac{\hbar}{64\pi^2}\Bigg(-\frac{\Lambda^4}{2}+\Lambda^2{\cal M}^2_i\nn
&&~~~~~~-\Big({\cal M}^2_i\Big)^2\ln\left(\frac{{\cal M}^2_i+\Lambda^2}{{\cal M}^2_i}\right)\Bigg)~.\nonumber
\eea
To compare this expression with Eq.(\ref{eq:Uad_final}), we follow the usual procedure, redefining the bare potential as
\bea
U_i&\to& U_i+\frac{\hbar}{64\pi^2}\left[\frac{\Lambda^4}{2}-\Lambda^2 {\cal M}^2_i\right.\\
&&~~~~~~~~~~~~~~~~~~\left.+\Big({\cal M}^2_i\Big)^2\left(\ln\left(\frac{\Lambda^2}{\mu^2}\right)+C\right)\right]~,\nonumber
\eea
where $\mu$ is a renormalisation scale and $C$ is a constant depending on the renormalisation scheme used, which is defined by the choice of finite terms. 
Keeping the first order in $\hbar$ only, the renormalised one-loop 1PI effective potential is then
\be \label{eq:U1ren01}
U^{(1)}_{ren}=U_i
+\frac{\hbar}{64\pi^2}\Big({\cal M}^2_i\Big)^2\left(\ln\left(\frac{{\cal M}^2_i}{\mu^2}\right)+C\right)~,
\ee
where the limit $\Lambda\to\infty$ is taken. The latter expression recovers the result (\ref{eq:Uad_final}), with $C=-3/2$.

\section{Wilsonian Cosmology}
\label{sec:wilsonian-cosmo}

From this point we simplify the notation for the potential and write $U_\kappa(\phi)=U$.

At any fixed times, modes that are within the Hubble horizon (UV, or {\em sub-horizon} modes) contribute to the scalar quantum field theory, whereas modes with wavelengths larger than the Hubble radius (IR, or {\em super-horizon} modes) live in the effective field theory in which sub-horizon modes have been integrated out. The Hubble scale $H(t)$ is thus a perfect candidate for our coarse-graining scale $\kappa(t)$, as it really corresponds to the scale at which the effective field theory of inflation is defined at any time during the inflation era. This picture corresponds precisely to the framework used in stochastic inflation. In this context, the coarse-graining is usually defined at a scale
\be
\label{eq:kappachoice}\kappa= H~,
\ee
such that the flow equation becomes
\be\label{flow2}
\ddot X+H\dot X+\left(H^2+\partial^2_\phi U-\frac{R}{6}\right)X=\frac{\hbar H^4}{32\pi^2}~,\nonumber
\ee
with 
\be
X\equiv H^{-1}\partial_H U~,
\ee
and where we consider a minimally coupled scalar ($\xi=0$).

Note that, depending on the context, Wilsonian approaches usually let the coarse-graining scale vary monotonously. Here, instead, $H(t)$ could in principle vary non-monotonously, as  the equations we use are reversible, so valid independently of the sign of $\dot H$.

The flow equation (\ref{flow2}) depends on a time-dependent background, whose evolution is provided by the field equations, that we derive here.
The action describing the homogeneous time-dependent inflaton is 
\bea
\Sigma&=&\int\! d^4\!x\sqrt{-g}\left(\frac{M_p^2}{2}R-\frac{1}{2}\partial_0\phi\partial^0\phi-U\right)\\
&=&\int\left\{\frac{3}{2}M_p^2\left[-\frac{2}{g_{00}}\left(\frac{\ddot a}{a}+\left(\frac{\dot a}{a}\right)^2\right)
+\frac{\dot g_{00}}{g_{00}^2}\frac{\dot a}{a}\right]\right.\nn
&&\left.~~~~~~-\frac{1}{2g_{00}}(\dot\phi)^2-U\right\}~\sqrt{-g_{00}}~a^3~d^4x~,\nonumber
\eea
and the field equations are obtained from the variation with respect to the dynamical variables $g_{00}, a, \phi$.
When taking the variation of the action with respect to $g_{00}$ before setting it to $-1$, we should take into account the covariant definition
of the Hubble rate with respect to time redefinition, which is
\be\label{defH}
H=\frac{1}{\sqrt{-g_{00}}}~\frac{\dot a}{a}~.
\ee
The dependence on $g_{00}$ is essential
to make $H$ a scalar under time redefinition, similarly to the inflaton (the expression (\ref{defH}) is proportional to the extrinsic curvature for FLRW spacetime).
As explained below, this is the consistent way to derive the correct Hamiltonian constraint and obtain the continuity equation.

The following functional derivatives leads to the required equations:
\begin{itemize}
\item $\delta \Sigma/\delta\phi=0$ for the scalar field evolution 
\be\label{dynphi}
\ddot\phi+3H\dot\phi+\partial_\phi U=0~,
\ee
\item $\delta \Sigma/\delta g_{00}=0$ for the Hamiltonian constraint
\be\label{constraint}
3M_p^2H^2=\rho=\frac{1}{2}(\dot\phi)^2+U - H^2X~,
\ee
where $\rho$ is the energy density;
\item $\delta \Sigma/\delta a=0$ for the scale factor dynamical equation  
\bea\label{dyna}
&&-M_p^2\left(2\frac{\ddot a}{a}+H^2\right)=p\\
&=&\frac{1}{2}(\dot\phi)^2-U + H^2X + \frac{1}{3}\frac{d(HX)}{dt}~,\nonumber
\eea
where $p$ is the pressure.
\end{itemize}
The last two equations lead to the dynamical equation for $H$
\be\label{dynH}
2M_p^2~\dot H=-(\dot\phi)^2-\frac{1}{3}\frac{d(HX)}{dt}~.
\ee
If one takes into account 
\be
\dot U=\dot H HX+\dot\phi \partial_\phi U~,
\ee
one can check that the two Friedmann equations (\ref{constraint}) and (\ref{dynH}) imply the scalar field equation (\ref{dynphi}), as expected, 
which is consistent with the continuity equation 
\be
\dot\rho=-3H(\rho+p)~.
\ee
Expressing all the equations derived above in terms of number of $e$-folds $N\equiv \ln(a)$ rather than cosmic time (primes denoting derivatives with respect to $N$), we obtain the following set of equations
\noindent\refstepcounter{equation}\label{eq:ODE}%
\begin{tcolorbox}[enhanced, colback=white, colframe=black, boxrule=0.5pt,
                  left=3pt, right=3pt, top=-4pt, bottom=2pt]
  \[
  \begin{aligned}
      X''&=\frac{\hbar H^2}{32\pi^2}-\left(1+\frac{H'}{H}\right)X'
           \\
           &-\left(\frac{\partial^2_\phi U}{H^2}-\frac{H'}{H}-1\right)X\,,\\
      \phi''&=-\left(3+\frac{H'}{H}\right)\phi'-\frac{\partial_\phi U}{H^2}\,,\\
      H'&=-\frac{1}{2M_p^2}\left[H(\phi')^2+\frac{1}{3}\left(HX\right)'\right]\,.
  \end{aligned}
  \tag*{\((\theequation)\)} 
  \]
\end{tcolorbox}

\section{Study of de Sitter regime}

Now that we have derived the equations of motion that encode the coupled dynamics of a single scalar field, its effective potential, and the Hubble scale, we are ready to explore the potential impact of this dynamic on specific potential examples and how it may impact cosmology.

In the following, a de Sitter configuration is defined by a constant Hubble rate, namely
\begin{equation}
    \epsilonH\equiv-\frac{H'}{H}=0\,.
\end{equation}
It is a fixed point of the cosmological dynamics only if the remaining evolution equations also vanish at the same configuration. The existence of such a point does not by itself imply stability. To determine whether de Sitter is stable, we perturb all the homogeneous dynamical variables slightly away from their fixed-point values and study their linearised evolution. If every perturbation decreases with the number of e-folds, nearby cosmological solutions return to constant $H$, and the de Sitter point is an attractor. If at least one perturbation grows, nearby solutions generically move away from constant $H$, and de Sitter is unstable; when some perturbations decay while another grows, the point is a saddle. 

We consider two cases for which the assumed form of the potential is preserved by the evolution. We first study a field-independent potential and then add a positive quadratic curvature while placing the scalar field at its minimum. In both cases the cosmological dynamics reduces to a two-dimensional first order system, allowing the constant-Hubble solutions and their stability to be characterised analytically.

\subsection{Flat Potential}
Let us start with the simplest possible case of a flat potential and investigate under which conditions a stable de Sitter solution exists for our set of flow and Friedmann equations. 

To do so, let us first restrict ourselves to the case where the field is at rest ($\phi'=0$) on a perfectly flat potential
\be U(\phi, H)=U_0(H)\,.\ee
With such initial conditions, the fact that the gradient and curvature of the potential vanish at the initial time,
\be
\partial_\phi U = \partial_\phi^2 U = 0\,,\ee
guarantees that the evolution of $X$ is also field-independent, so that the potential remains flat while it is running. Under these conditions, the equations of motion of Eqs.~\eqref{eq:ODE} become
\noindent\refstepcounter{equation}\label{eq:ODE_Lambda}%
\begin{tcolorbox}[enhanced, colback=white, colframe=black, boxrule=0.5pt,
                  left=3pt, right=3pt, top=-4pt, bottom=2pt]
  \[
  \hspace{-2pt}\textbf{(Flat Case)\qquad}\begin{aligned}
      X''&=\frac{\hbar H^2}{32\pi^2}-\left(1+\frac{H'}{H}\right)X'
           \\
           &~~~~+\left(1+\frac{H'}{H}\right)X\,,\\
      \frac{H'}{H}&=-\frac{1}{6M_p^2}\frac{X'}{1+X/6M_p^2}\,.
  \end{aligned}
  \tag*{\((\theequation)\)} 
  \]
\end{tcolorbox}

\vspace{10pt}
\noindent
{\bf Running de Sitter.\ }
From these equations, it is clear that if one decides to forget about the initial flow of the cosmological constant, by setting initial conditions such that 
\be U(\phi,H_{\rm in}) = U_0(H_{\rm in})\,,\ H=H_{\rm in}\,,\ H'=0\,,\ee
but enforcing that $X_{\rm in}=0$ and $X_{\rm in}'=0$ at the initial time, then the cosmological constant has no alternative but to run. 

The question is then whether the system can remain close to de Sitter, or whether the flow equation dynamically drives it away from the de Sitter solution. This can be understood analytically at early times by scrutinising the evolution of the system in the vicinity of its nearly de Sitter initial conditions, where
\be 
H\approx H_{\rm in}\,,\qquad \frac{H'}{H}\approx 0\,, 
\ee 
such that the first equation in Eq.~\eqref{eq:ODE} reduces to
\be
X''+X'-X\approx\frac{\hbar H_{\rm in}^2}{32\pi^2}\,.
\ee
Introducing the eigenvalues 
\be 
r_\pm = \frac{-1\pm\sqrt{5}}{2}\,, 
\ee 
and as an initial condition that $X=X'=0$, the solution of this equation is 
\bea 
X \approx \frac{-\hbar H_{\rm in}^2}{32\pi^2}\left[1 + \frac{r_-}{\sqrt{5}} e^{r_+N} -\frac{r_+}{\sqrt{5}} e^{r_-N}\right]\,.\nonumber\\\label{eq:Xsol}
\eea 
This expression makes explicit why \(X=0\) is not a fixed point of the flow equation: the source term proportional to \(\hbar H_{\rm in}^2\) excites the growing homogeneous mode. After a few $e$-folds, this mode dominates Eq.~\eqref{eq:Xsol}, and one can estimate the number of $e$-folds it takes for the running of the cosmological constant to induce sizeable changes to the initial condition, by defining
\be\left.\frac{H^2 X}{U}\right|_{N=N_c}\equiv 1\,,\ee
obtaining
\bea
N_c&\simeq& \frac{1}{r_+} \ln\left[ 96\pi^2 \frac{\sqrt{5}}{-r_-}  \frac{M_p^2}{\hbar H_{\rm in}^2}\right]\,, \nonumber\\
&\approx & 11.6 + 1.62 \ln\left[\frac{M_p^2}{\hbar H_{\rm in}^2} \right] \,.
\eea
Interestingly, this means that a cosmological constant with a very small value can survive for many $e$-folds before it is destabilised by the running of the cosmological constant due to a massless scalar, as using the present value of Hubble ($H_{\rm in}\sim 6\times 10^{-61}M_p$) leads to $N_c\sim 460$.

\vspace{10pt}
\noindent
{\bf Hubble-Controlled de Sitter.\ }
Let us now allow the flow of the potential, $X$, to be non zero at initial time, and explore whether an exact de Sitter solution can be protected by the dynamics. At a constant-Hubble fixed point, both $H_\star$ and $X_\star$ are independent of $N$, so that  $ H'=X'=X''=0$. The first equation thus simply reduces to
\be\label{eq:FlowCC}
X_{\star}=-\frac{\hbar H_\star^2}{32\pi^2}~.
\ee
The two Friedmann equations we obtained in Sec.~\ref{sec:wilsonian-cosmo} are then equivalent to 
\be
H_\star^2 X_{\star} = U_0-3M_p^2H_\star^2~,
\ee
such that 
\be\label{relationUHT_1}
U_0=3M_p^2H_\star^2-\frac{ \hbar H_\star^4}{32\pi^2}~.
\ee
This case is trivial, but representative of what happens physically: for a given value of the cosmological constant $U_0$, the Hubble scale is given by Eq.~\eqref{relationUHT_1}, and one needs to fine-tune the flow of the cosmological constant with the Hubble scale ($X_\star\not =0$) using Eq.~\eqref{eq:FlowCC} such that all the time derivatives in the master equation are zero and $H$ can remain constant. In this sense, a pure de Sitter solution cannot exist without being assisted by an appropriate RG flow. The existence of this solution does not imply stability: we must still determine whether nearby solutions approach it or move away.

\vspace{10pt}
\noindent
{\bf Phase Portrait.\ } 
In order to study the overall dynamics of the system, it is instructive to press the equations of motion as much as possible to reduce the dimensionality of the system. In the flat-potential limit, and for vanishing initial field velocity, the scalar equation reads
\be
\phi''=-\left(3+\frac{H'}{H}\right)\phi'\,.
\ee
Therefore, if the field is initially at rest, it remains at rest, and the only non-trivial evolution is that of \(X\) and \(H\), whose dynamic is driven by the equations
\be
X''
=
\frac{ \hbar H^2}{32\pi^2}
-\left(1+\frac{H'}{H}\right)X'
+\left(1+\frac{H'}{H}\right)X\,,
\label{eq:flat_X_second_order}
\ee
and
\be
\frac{H'}{H}
=
-\frac{1}{6M_p^2}
\frac{X'}{1+X/(6M_p^2)}\,.
\label{eq:flat_H_X_relation}
\ee
Fortunately, the latter equation can be expressed as a total logarithmic derivative,
\be\label{eq:H0def}
H\left(1+\frac{X}{6M_p^2}\right)
=H_0\,,
\ee
where $H_0$ is a constant, which can be interpreted as the value of the Hubble constant in the limit of vanishing flow $X=0$. This relation allows us to eliminate \(X\) from the flat-potential dynamics. To simplify the analysis, it is then convenient to introduce the dimensionless variables
\be
h\equiv \frac{H}{H_0}\,,
\qquad
\epsilonH\equiv -\frac{H'}{H}\,.
\ee
and using Eq.~\eqref{eq:flat_X_second_order}, we obtain the closed first-order system
\be
\begin{aligned}
h' &= -\epsilonH h\,,
\\[2mm]
\epsilonH'
&=
\gamma h^3
-(2-h)\epsilonH
+
(1-h)\,,
\end{aligned}
\label{eq:xh_system_flat_C}
\ee
where
\be
\gamma\equiv
\frac{\hbar H_0^2}{192\pi^2M_p^2}\,.
\ee
Under this form, the stability of the system becomes relatively trivial to study. Looking for an equilibrium $(h_\star,\epsilonH_{\star})$  such that $h'=\epsilonH_{}'=0$, one obtains two distinct equilibria. The first one is located at
\be
\epsilonH_{\star} = 0\,,\qquad 
1-h_\star
+
\gamma h_\star^3
=0\,.
\label{eq:assisted_dS_phase_point}
\ee
Note that although this system leads to several solutions for $h_\star$, all solutions satisfy the Hubble-controlled de Sitter condition of Eq.~\eqref{eq:FlowCC} and the fact that there exist several roots only comes from the fact that the equation is expressed in terms of the constant $H_0$ rather than in terms of the actual Hubble constant $H_\star$. However, the only solution that satisfies $H_\star\ll M_p$ (or equivalently $h_\star H_0 \ll M_p$), corresponds to a solution $|1-h_\star|\ll 1$, leading to the only solution
\be
h_{\star}
=
1+\gamma+\mathcal O(\gamma^2)\,.
\ee
Linearising the system around this equilibrium, and studying the eigenvalues of its Jacobian
\be
\lambda_{{\rm dS},\pm}
=
\frac{-1\pm
\sqrt{
5
}}{2}
+\mathcal O(\gamma)\,,
\ee
reveals that the Hubble-controlled de Sitter equilibrium is a saddle point.

The reduced system also possesses the asymptotic limit
\be
(h,\epsilonH)_{\rm Mink.}=\left(0\,,\ \frac{1}{2}\right)\,,
\ee
corresponding to the Minkowski limit in which $h\to 0$. Note that technically, for $h=0$,  $\epsilonH$ is not properly defined, so this second equilibrium actually corresponds to an asymptotic limit.
The eigenvalues of the linearised system  in this limit are both negative,
meaning that this asymptote is a local attractor trajectory.  This attractor corresponds to a scaling solution along which
\be
\epsilonH=\frac{1}{2}\,,\quad H\propto \exp\left(-\frac{1}{2}N\right)\,.\ee

Note that due to Eq.~\eqref{eq:H0def}, one can obtain a relation between the value of the potential $U$ and the value of $h$, 
\be
U_0=3 M_p^2 H_0^2 h(2-h)\,,\ee
such that in the region of the plane $(h,\epsilonH)$ where $h>2$, it is clear that $U=U_0<0$. 
\begin{figure*}
    \centering
    \includegraphics[width=0.8\linewidth]{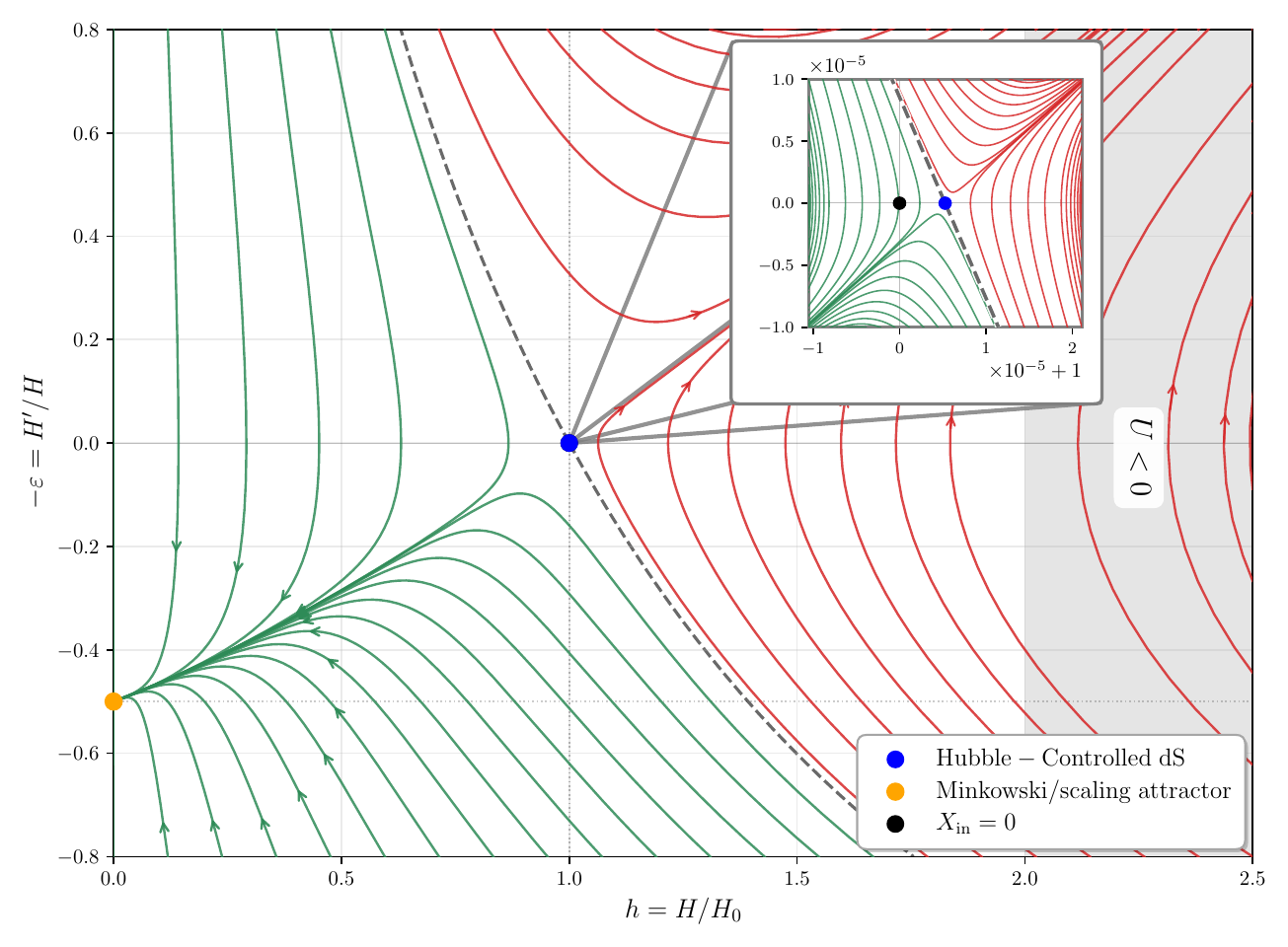}
    \caption{\label{fig:massless} \footnotesize
Phase portrait of the massless system in the
\((h,\epsilonH)\) plane for \(H_0=0.1M_p\).
The Hubble-controlled de Sitter equilibrium (blue dot) is a saddle, while the Minkowski
scaling solution (orange dot) is an asymptotic attractor. Green trajectories belong
to the Minkowski basin of attraction, whereas red trajectories lie on
the runaway side of the stable-manifold separatrix. The gray-shaded region
\(h>2\) corresponds to \(U<0\).
}
\end{figure*}

This dynamics is depicted in Fig.\ref{fig:massless} in the case where $H_0=0.1 M_p$. Trajectories are depicted in green when they are attracted towards the
scaling solution. Red trajectories lie on its runaway side and evolve towards the
pathological asymptotic regime
\be
H\to\infty\,,
\qquad
\epsilonH\to\infty\,,
\qquad
U\to-\infty\,,
\ee
which leads outside the semi-classical regime. Finally, the gray dashed curve is the stable hyperspace of the Hubble-controlled de Sitter
saddle, acting as a separatrix between the attractor basin and the runaway patch. Notably, the running de Sitter solution studied above, starting with no initial running ($X_{\rm in}=0$, or equivalently $h_{\rm in}=1$),  starts outside the runaway region.  

\subsection{Quadratic Potential}
We now investigate how the previous dynamics is modified when the scalar
potential acquires a non-vanishing curvature. Allowing initially for a
possible RG dependence of the scalar-field mass term, we write
\be
U(\phi,H)=U_0(H)+\frac{1}{2}m^2(H)\phi^2\,.
\ee
Since the flow of the potential depends on the background field only
through \(\partial_\phi^2U=m^2(H)\), which is independent of \(\phi\), the
flow \(X\) is itself field-independent. Consequently,
\be
\partial_H m^2 = \partial_H\partial_\phi^2U = \partial_\phi^2 (\partial_H U) \propto \partial_\phi^2X
=
H^{-1}\frac{dm^2}{dH}
=
0\,,
\ee
and the mass is constant along the RG flow, $ m^2(H)=m^2$. We therefore consider
\be
U(\phi,H)=U_0(H)+\frac{1}{2}m^2\phi^2\,,
\qquad m^2>0\,,
\ee
and initially place the scalar at the minimum of its potential,
\be
\phi=\phi'=0\,.
\ee
Since
\be
\left.\partial_\phi U\right|_{\phi=0}=0\,,
\qquad
\partial_\phi^2U=m^2\,,
\ee
the hypersurface \(\phi=\phi'=0\) remains invariant under the evolution. The non-vanishing curvature of the potential nevertheless modifies the flow equation for \(X\), and the equations of motion become
\noindent\refstepcounter{equation}\label{eq:ODE_massive}%
\begin{tcolorbox}[enhanced, colback=white, colframe=black, boxrule=0.5pt,
                  left=3pt, right=3pt, top=-4pt, bottom=2pt]
  \[
  \hspace{-2pt}\textbf{(Massive Case)\qquad}
  \begin{aligned}
      X''
      &=
      \frac{\hbar H^2}{32\pi^2}
      -\left(1+\frac{H'}{H}\right)X'
      \\
      &\hspace{-5pt}
      +\left(
      1+\frac{H'}{H}
      -\frac{m^2}{H^2}
      \right)X\,,
      \\[1mm]
      \frac{H'}{H}
      &=
      -\frac{1}{6M_p^2}
      \frac{X'}{1+X/(6M_p^2)}\,.
  \end{aligned}
  \tag*{\((\theequation)\)}
  \]
\end{tcolorbox}

\vspace{10pt}
\noindent
{\bf Running de Sitter.\ }
The local effect of the mass term can first be understood by considering
initial conditions close to an un-flowing de Sitter solution,
\be
H\approx H_{\rm in}\,,
\qquad
\frac{H'}{H}\approx0\,,
\qquad
X=X'=0\,.
\ee
The equation for \(X\) then reduces to
\be
X''+X'-\left(1-\frac{m^2}{H_{\rm in}^2}\right)X
\simeq
\frac{\hbar H_{\rm in}^2}{32\pi^2}\,.
\label{eq:massive_local_X}
\ee
Introducing
\be
\Delta_m\equiv
\sqrt{
5-\frac{4m^2}{H_{\rm in}^2}
}\,,
\qquad
r_\pm=\frac{-1\pm\Delta_m}{2}\,,
\ee
one obtains, for \(H_{\rm in}\neq m\),
\be
X\simeq
-\frac{\hbar H_{\rm in}^2}
{32\pi^2(1-m^2/H_{\rm in}^2)}
\left[
1+\frac{r_-}{\Delta_m}e^{r_+N}
-\frac{r_+}{\Delta_m}e^{r_-N}
\right].
\label{eq:Xsol_massive}
\ee
The ordering between $H_{\rm in}$ and $m$ determines the local behaviour. If
\be
H_{\rm in}>m\,,
\ee
then the eigenvalue \(r_+\) is positive.
The growing mode responsible for destabilising the flat-potential de
Sitter solution therefore remains present. Conversely, if
\be
H_{\rm in}<m\,,
\ee
both homogeneous modes have negative real
parts. The mass term then removes the growing flow mode and locally
stabilises the de Sitter configuration.
Exactly at \(H_{\rm in}=m\), the
solution with \(X=X'=0\) initially behaves as
\be
X\simeq
\frac{\hbar H_{\rm in}^2}{32\pi^2}
\left(N-1+e^{-N}\right)\,.
\ee
The transition at \(H=m\) is therefore a limiting case at the linear level.

\vspace{10pt}
\noindent
{\bf Hubble-Controlled vs Mass-Controlled de Sitter.\ }
A constant-Hubble solution again requires
\be
H'=X'=X''=0\,.
\ee
The flow equation then imposes
\be
X_\star
=
-\frac{\hbar H_\star^2}
{32\pi^2\left(
1-m^2/H_\star^2
\right)}\,.
\label{eq:massive_assisted_flow}
\ee
The corresponding Friedmann equation gives
\be
U_0
=
3M_p^2H_\star^2+H_\star^2X_\star\,,
\ee
or equivalently
\be
U_0
=
3M_p^2H_\star^2
-
\frac{\hbar H_\star^4}
{32\pi^2\left(
1-m^2/H_\star^2
\right)}\,.
\label{eq:massive_assisted_potential}
\ee
The mass term therefore modifies the RG flow required to support a
constant Hubble scale. In particular, the response of \(X\) becomes large
when \(H_\star\) approaches $m$\,. 
\label{eq:Hm_definition_massive}
Although this limit looks pathological at first sight, we will see that the dynamics remains regular in this region. More importantly, the introduction of a non-zero mass term breaks the one-to-one relation that existed in Eq.~\eqref{eq:FlowCC} in the massless case. This means that there exist two different de Sitter fixed points. As we will see, one is analogous to the Hubble-controlled de Sitter point found in the massless case, and the other one is controlled by the apparition of the new mass term.

\vspace{10pt}
\noindent
{\bf Phase Portrait.\ }
The last equation of Eq.~\eqref{eq:ODE_massive} is unchanged by the mass
term and can again be integrated to give
\be
H\left(1+\frac{X}{6M_p^2}\right)=H_0\,.
\label{eq:H0def_massive}
\ee
We can thus define the same set of dimensionless variables
\be
h\equiv\frac{H}{H_0}\,,
\qquad
\epsilonH\equiv-\frac{H'}{H}\,,
\ee
together with
\be
\gamma\equiv
\frac{\hbar H_0^2}{192\pi^2M_p^2}\,,\qquad \mu\equiv\frac{m}{H_0}\,.
\ee
With those notations, the massive dynamics now reduces to
\be
\begin{aligned}
h'
&=
-\epsilonH h\,,
\\[2mm]
\epsilonH'
&=
\gamma h^3
-(2-h)\epsilonH
+
\left(
1-\frac{\mu^2}{h^2}
\right)(1-h)\,,
\end{aligned}
\label{eq:h_epsilon_massive_system}
\ee
where one can easily check that the massless system is recovered by setting \(\mu \to 0\).

Armed with these equations, we can now search for the equilibrium points of this system. An interior ($h>0$) equilibrium has \(\epsilonH_{\star}=0\) and satisfies
\be
\gamma h_\star^3
+
\left(
1-\frac{\mu^2}{h_\star^2}
\right)(1-h_\star)
=0\,.
\label{eq:massive_fixed_equation}
\ee
Within the expanding ($h>0$), sub-Planckian domain ($h_\star H_0<M_p$), this equation possesses two
relevant solutions. In the limit \(H_0\ll M_p\), at zeroth order in $\gamma$, they reduce to $h_\star=1$, the Hubble-controlled de Sitter solution, and $h_\star=\mu$, a new mass-controlled equilibrium. 
Away from the degenerate case
\(m=H_0\) (corresponding to $\mu = 1$), the source term shifts these values by a small amount,
\be
h_{\rm dS}
=
1+
\frac{\gamma}
{1-\mu^2}
+\mathcal O(\gamma^2)\,,
\label{eq:hds_massive_expansion}
\ee
and
\be
h_{m}
=
\mu
-
\frac{\gamma \mu^4}
{2(1-\mu)}
+\mathcal O(\gamma^2)\,.
\label{eq:hm_massive_expansion}
\ee
Naturally, these expansions cease to be valid when \(\mu\simeq1\), where the two
classical equilibria become degenerate.

The critical case must be analysed directly. Setting $\mu=1$ in Eq. \eqref{eq:massive_fixed_equation} gives 
\begin{equation}
    \gamma h_\star^3- \frac{(h_\star-1)^2(h_\star+1)}{h_\star^2}=0\, ,
\end{equation}
For $\gamma\ll 1$, the two solutions near $h_\star=1$ are $h = 1 \pm \sqrt{{\gamma}/{2}} + \mathcal{O(\gamma)}$.
Thus, at finite $\gamma$, there is no fixed point exactly at
$H=m=H_0$; instead, there are two nearby constant-Hubble solutions on
either side of this value.

Linearising the system around these equilibria, one can compute the Jacobian of the dynamical system and study the respective stability of the two points. The eigenvalues of this Jacobian are given by
\be
\lambda_{{\rm dS},\pm}
=
\frac{-1\pm
\sqrt{
1+4(1-\mu^2)
}}{2}
+\mathcal O(\gamma)\,,
\ee
for the Hubble-controlled de Sitter point, and
\be
\lambda_{m,\pm}
=
\frac{
\mu-2
\pm
\sqrt{
(\mu-2)^2
-8(1-\mu)
}
}{2}
+\mathcal O(\gamma)\,,
\ee
\begin{figure}
    \centering
\includegraphics[width=\linewidth]{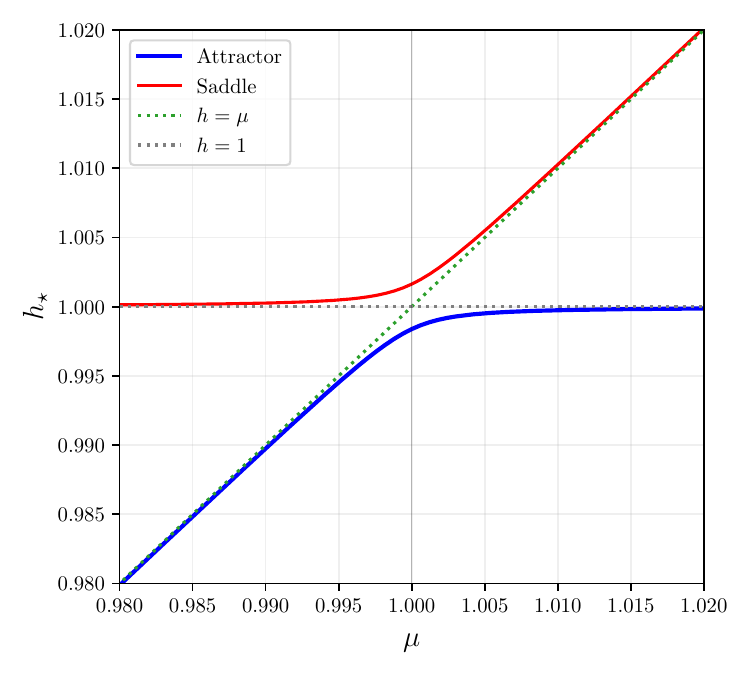}
    \caption{\label{fig:swap} Evolution of the two sub-Planckian roots of Eq.~ \eqref{eq:massive_fixed_equation}, $h_\star$, as a function of $\mu$, transitioning from the IR regime ($\mu<1$) to the UV regime ($\mu>1$). Reference values of $h_{\rm dS}\approx 1$ and $h_{m}\approx \mu$ are indicated as dotted lines for reference.}
\end{figure}
for the mass-controlled equilibrium. Remarkably, as can be seen on Fig.~\ref{fig:swap}, depending on the sign of $1-\mu$, one of the eigenvalues flips sign in both cases, meaning that the two equilibrium points flip stability: For \(\mu<1\), the Hubble-controlled de Sitter point
is a saddle point and the mass-controlled point is an attractor. In the opposite case ($\mu>1$), the Hubble-controlled de Sitter point becomes an attractor and the mass-controlled point becomes a saddle. 

\begin{figure}
    \centering
\includegraphics[width=\linewidth]{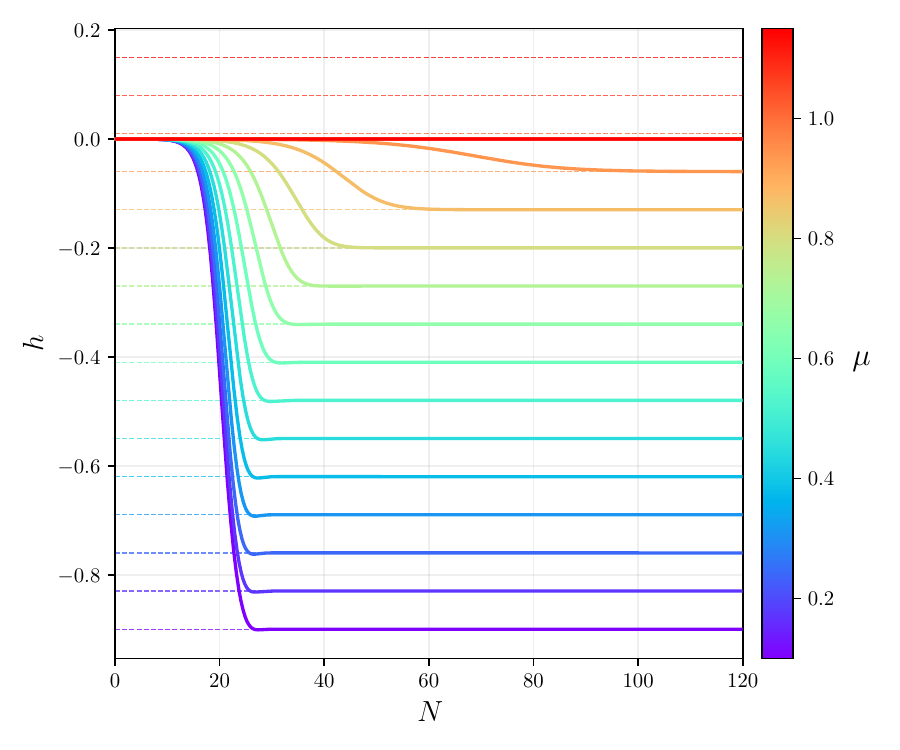}
    \caption{\label{fig:H_IR} Evolution of $h$ as a function of $N$ for various values of the mass ranging from the IR regime where $\mu\equiv m/H_0 \ll 1$ (purple colour) to the UV regime where $\mu\gg 1$ (red colour). Individual values of $\mu$ are indicated by horizontal dashed lines.}
\end{figure}

\begin{figure}
    \centering
\includegraphics[width=\linewidth]{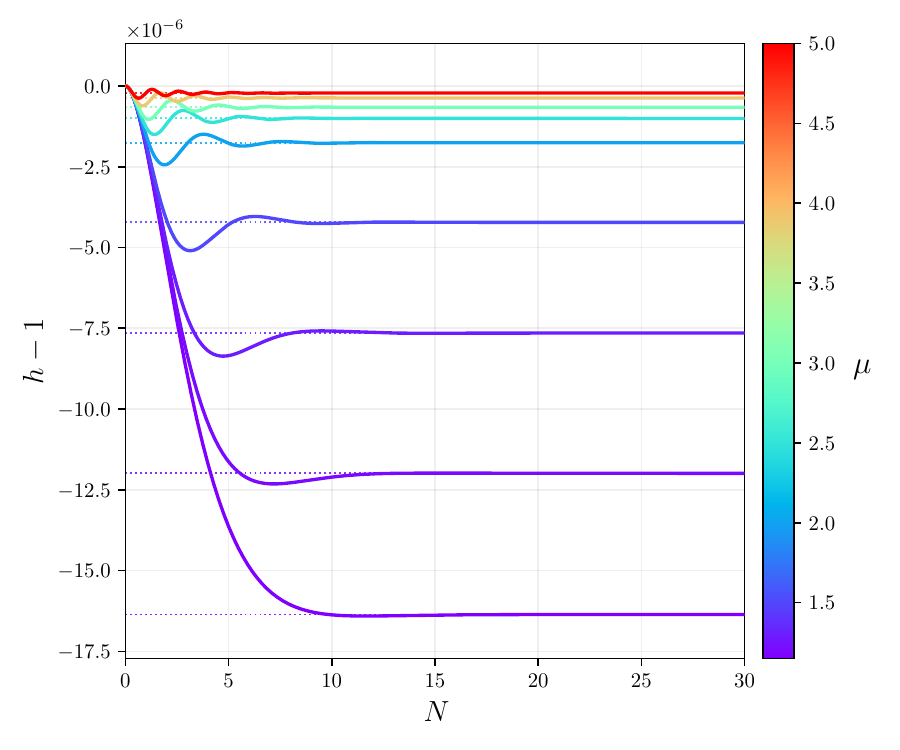}
    \caption{\label{fig:H_UV} Evolution of $h-1$ as a function of $N$ for various values of the mass within the UV regime where $\mu\gg 1$. Individual values of $h_{\rm dS}-1\approx \gamma/(1-\mu^2)$ are indicated by horizontal dotted lines.}
\end{figure}

We note that, in the context of Wilsonian studies, the freezing of flows when the running scale $H$ becomes small compared to the mass $m$ is a generic feature.
Indeed, the observational scale $H^{-1}$ is then large compared to the Compton wavelength $m^{-1}$ of the particle, and the latter
therefore behaves classically, suppressing Wilsonian flows. This transition is illustrated in Fig.~\ref{fig:H_IR}, in which $\mu$ is varied from values $\ll 1$ to values $\gtrsim 1$ and the evolution of $h$ is compared to the mass-controlled de Sitter value $h_{m}$ (dashed line). One can see that in the IR limit, the mass-controlled de Sitter configuration is indeed an attractor, where it stops attracting the dynamics for $\mu\gtrsim 1$. In Fig.~\ref{fig:H_UV}, this is visualised further by zooming in around $h\approx 1$ for larger values of $\mu >1$ for which one can see that $h$ is attracted instead to the Hubble-controlled de Sitter attractor $h_{\rm dS}$ (dotted lines).

Depending on the parameters, the attractor can be either a stable node (real and negative eigenvalues)
or a stable focus (complex eigenvalues with negative real parts). Therefore, in the $\gamma\ll 1$ regime, the
dynamics is generically attracted towards the lower of the two
characteristic scales $H_0$ and $m$.

As in the flat-potential case, the Friedmann equation gives
\be
U_0
=
3M_p^2H_0^2h(2-h)
\ee
when \(\phi=0\). Hence
\be
h>2
\quad\Leftrightarrow\quad
U_0<0\,.
\ee

\begin{figure*}
    \centering
    \includegraphics[width=0.8\linewidth]
    {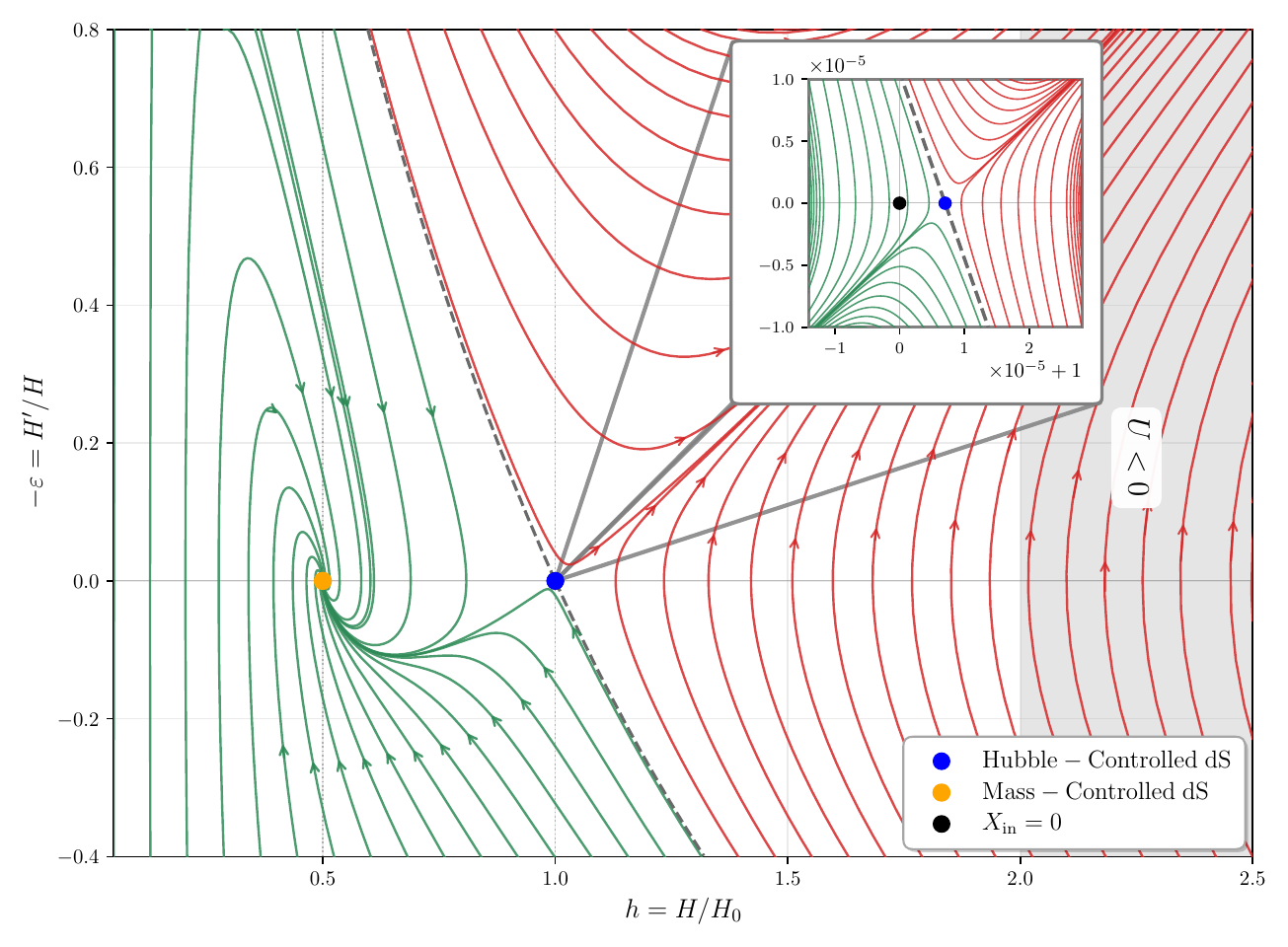}
    \caption{\label{fig:massive}
    Phase portrait of the massive system in the
    \((h,\epsilonH)\) plane for \(H_0=0.1M_p\), and \(m/H_0=0.5\).
    The Hubble-controlled de Sitter equilibrium is a saddle, while the
    mass-controlled equilibrium is a stable focus. Green trajectories
    belong to its basin of attraction, whereas red trajectories lie on
    the runaway side of the stable-manifold separatrix. The shaded region
    \(h>2\) corresponds to \(U_0<0\).
    }
    
\end{figure*}

\begin{figure*}
    \centering
    \includegraphics[width=0.8\linewidth]
    {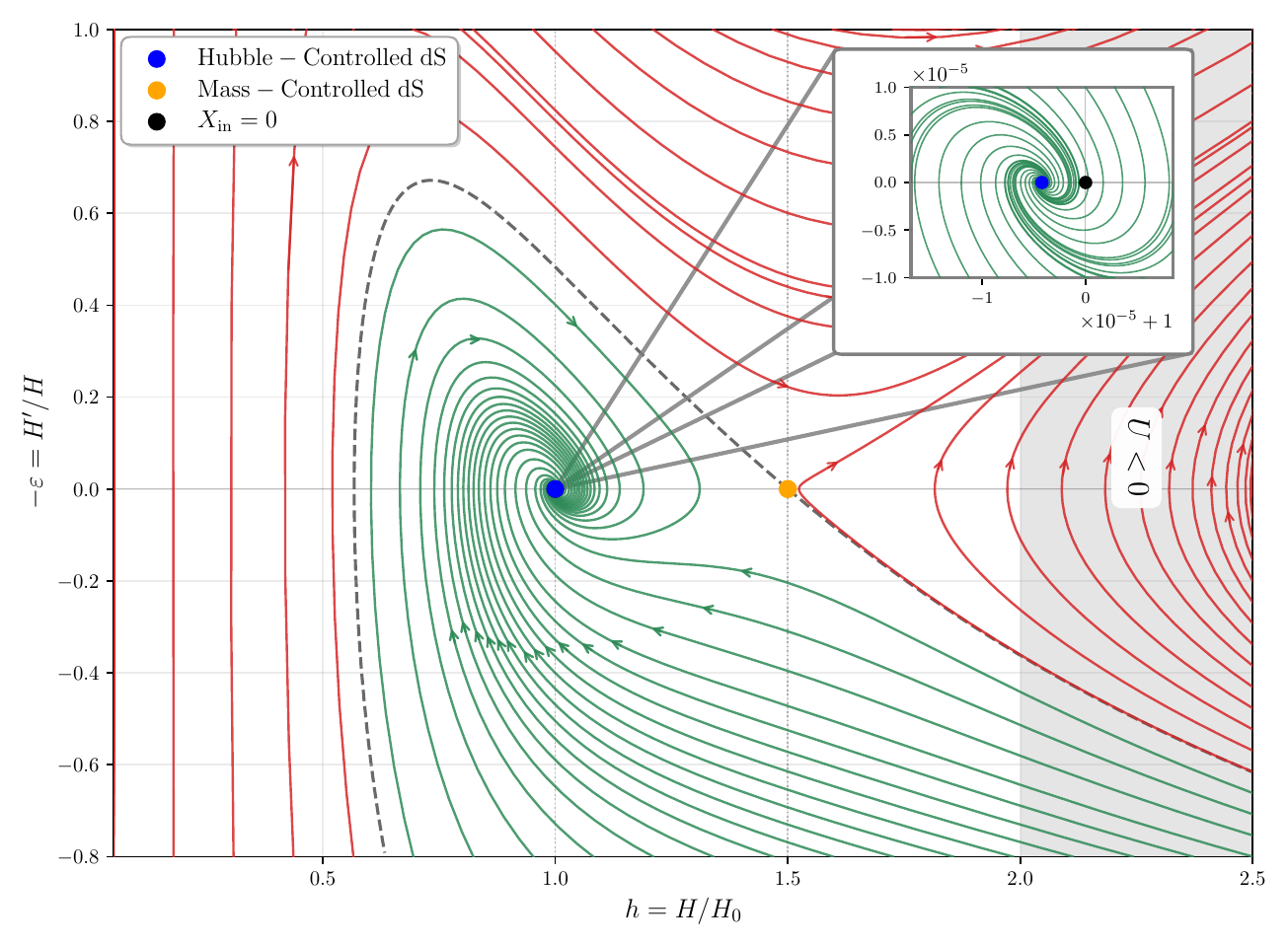}
    \caption{\label{fig:massive_m1p5}
    Same as FIG.~\ref{fig:massive}, but for $m/H_0=1.5$, and with the difference that the Hubble-controlled de Sitter point is now an attractor, while the mass-controlled equilibrium is a saddle.
    }
    
\end{figure*}

For the benchmark parameter
$H_0=0.1M_p$, 
the dynamics is illustrated in Fig.~\ref{fig:massive} and \ref{fig:massive_m1p5} in the cases where $m/H_0=0.5$ and $1.5$, respectively.
In both cases, the Minkowski scaling attractor that was present in the massless case is removed from the regular fixed-\(H_0\)
phase portrait. Indeed, the massive equations contain a contribution
proportional to \(m^2/H^2\), which diverges as \(H\to0\). Depending on the ratio
\(\mu=m/H_0\), the late-time attractor is instead either
the mass-controlled equilibrium or the Hubble-controlled de Sitter equilibrium.
Note that for \(m\ll H_0\), however, the mass term is initially negligible and trajectories
may transiently follow the massless Minkowski scaling solution before
being diverted towards the relevant finite-Hubble attractor. 

Another important aspect of the $m>H_0$ case is that, unlike in the $m<H_0$ case, the basin of attraction of $H_0$ does not span all the way to $h\to 0$. Instead, curves starting with a very small $h$ may quickly grow to relatively large $H'/H$, circumventing the basin of attraction, and quickly running towards the $U<0$ region. Again, we note that the running de Sitter solution studied above, starting with no initial running ($X_{\rm in}=0$, or equivalently $h_{\rm in}=1$),  starts outside the runaway region.

\subsection{Stability of the $\phi'=0$ Regime}
\label{sec:velocity_stability}

So far, we restricted our analysis to the case $\phi'=0$, with $\phi=0$ in the quadratic case. We now examine whether this restriction is stable
against a small initial scalar velocity.

Let \((\bar H,\bar X)\) denote any of the background solutions studied
above and introduce a small scalar perturbation of amplitude
\(\delta \phi\), such that
\be
\phi=\bar\phi+\delta\,\phi\,,\ee
where \(\bar\phi\) may be any constant for a flat potential, whereas
\(\bar\phi=0\) in the quadratic case. 

Remarkably, such a perturbation enters in the last line of Eq.~\eqref{eq:ODE} only through $(\phi')^2$, while $\partial_\phi^2 U$ is field-independent for the two potentials considered, such that perturbations to the dynamics of $H$ (and therefore the dynamics of $X$) are second order. This means that, at first order in perturbation theory, the evolution of $\delta\phi$ can be studied as evolving in the \((\bar H,\bar X)\) background. The scalar perturbation
therefore decouples from the background dynamics at linear order and
satisfies
\be
\delta\phi''
+
\left(
3+\frac{\bar H'}{\bar H}
\right)\delta\phi'
+
\frac{m^2}{\bar H^2}\delta\phi
=0\,,
\label{eq:linear_scalar_velocity}
\ee
where \(m=0\) in the flat-potential case.

\paragraph*{Flat potential.}
For \(m=0\), Eq.~\eqref{eq:linear_scalar_velocity} can be integrated
exactly, without assuming that the background Hubble rate is constant:
\be
\delta\phi'(N)
=
\delta\phi'_i\,
\frac{\bar H_i}{\bar H(N)}
e^{-3(N-N_i)}\,.
\ee
Any small initial kinetic energy is therefore rapidly diluted by cosmic expansion. 

\paragraph*{Quadratic potential.}
For a quadratic potential, it is useful to linearise the system around a constant-Hubble
equilibrium \(H=H_\star\) gives
\be
\delta\phi''+3\delta\phi'
+\frac{m^2}{H_\star^2}\delta\phi=0\,.
\label{eq:scalar_quadratic_linear}
\ee
The corresponding eigenvalues are
\be
s_\pm
=
\frac{-3\pm
\sqrt{9-4m^2/H_\star^2}}{2}\,,
\label{eq:scalar_eigenvalues}
\ee
whose real parts are negative for all \(m^2>0\). The equilibria are therefore stable in the scalar direction, independently of their stability within the reduced \((h,\epsilonH)\) phase plane. For
\(m/H_\star<3/2\), the scalar relaxes towards the minimum through an
overdamped evolution, while for \(m/H_\star>3/2\) it undergoes damped
oscillations with an envelope proportional to \(e^{-3N/2}\), corresponding to a very fast damping in a quasi-de Sitter universe.

\paragraph*{Shift of the invariant leaf.}
Although the scalar perturbation does not modify the background at
linear order, its kinetic energy produces a finite backreaction at second order in
\(\delta\phi\). In the static field case, the Friedmann equation could
be integrated to give
\be
H\left(1+\frac{X}{6M_p^2}\right)=H_0\,.
\ee
When \(\phi'\neq0\), the same combination instead obeys the exact
relation
\be
\left[
H\left(1+\frac{X}{6M_p^2}\right)
\right]'
=
-\frac{H}{2M_p^2}(\phi')^2
\leq0\,.
\label{eq:H0_drift}
\ee
Therefore, the integration constant $H_0$ that labels the $\phi'=0$ trajectories is no longer conserved. It is thus useful to define the instantaneous leaf parameter
\be
H_0(N)
\equiv
H(N)\left(1+\frac{X(N)}{6M_p^2}\right).
\ee
Integrating Eq.~\eqref{eq:H0_drift} gives
\be
H_0(N)
=
H_{0,i}
-
\frac{1}{2M_p^2}
\int_{N_i}^{N}
d\widetilde N\,
H(\widetilde N)
\bigl[\phi'(\widetilde N)\bigr]^2\,.
\label{eq:H0_integrated_shift}
\ee
Since the scalar velocity is damped, the integral converges and
\(H_0\) approaches a new constant,
\be
H_{0,\mathrm{f}}
=
H_{0,\mathrm{i}}
-
\frac{1}{2M_p^2}
\int_{N_i}^{\infty}
dN\,
H(\phi')^2\,.
\label{eq:H0_final}
\ee
A small velocity perturbation therefore does not invalidate the reduced
phase-space description. It produces a transient second-order
backreaction and eventually places the system on a neighbouring
static-field trajectory characterised by the slightly shifted
integration constant \(H_{0,\mathrm{f}}\).

In the flat-potential case, for an approximately constant Hubble rate and an initial kinetic scalar perturbation $\phi'(N_i)\equiv\phi_i'$, one has $\phi'(N)=\phi_i'e^{-3(N-N_i)}$,  and Eq.~\eqref{eq:H0_final} gives the estimate
\be
\frac{H_{0,\mathrm{f}}-H_{0,\mathrm{i}}}
     {H_{0,\mathrm{i}}}
\simeq
-\frac{(\phi'_i)^2}{12M_p^2}
\,.
\label{eq:H0_fractional_shift}
\ee
The shift is thus parametrically of the same order as the initial
kinetic-energy fraction.

We conclude that the field-at-rest sector is transversely stable. Small
initial scalar velocities are damped by the cosmological expansion, their
backreaction begins only at quadratic order, and the late-time evolution
is governed by the same reduced system studied above, with a
perturbatively shifted value of \(H_0\).

\section{Discussion and Conclusion}

{
In this paper, we have laid the foundation of a non-perturbative Wilsonian framework to describe the evolution of a scalar field theory in cosmology, accounting for the relevant quantum effects throughout cosmic evolution. Identifying the Hubble scale as the relevant cutoff discriminating between UV and IR modes in our quantum field theory, and integrating out UV modes dynamically, we derived a flow equation for the effective scalar potential as a function of time. Accounting for this flow in the Einstein equations, we obtained a set of quantum-corrected Friedmann equations, which allowed us to test the stability of de Sitter universes in two situations for which we could solve our equations exactly, without any truncation of the effective potential: the case where a scalar field is at rest in a flat and in a quadratic potential.}

{In both cases, we observed that a universe starting with a constant Hubble parameter and null running initially does not remain exactly at this initial de Sitter configuration, since the flow equation dynamically generates a non-vanishing running. The subsequent evolution, however, depends on the scalar potential.} 

{In the case of a pure cosmological constant (flat potential), we found that the induced running drives the universe towards  a scaling solution with $H^2\propto e^{-N}$. We also showed that it is possible to fine-tune the value of the cosmological constant running from the start so that de Sitter is an exact solution of the dynamics, but that this equilibrium is a saddle point, sitting on the edge of a dangerous region where Hubble would be driven to infinity in a short amount of time.}

In the case of a quadratic potential, we showed that the presence of the scalar field mass $m$ introduces a de Sitter-like attractor in the theory, located at $H \approx \min (m, H_0)$, 
where $H_0$ coincides with the initial Hubble parameter for $X_{\text{in}}=0$. In other words, and consistently with generic Wilsonian features, in the presence of a scalar field with $m\ll H_0$, the Hubble scale is attracted to $H\approx m$, whereas if $m\gg H_0$ from the start, the mass term basically stabilises the Universe in a de Sitter configuration very similar to its initial state $H\approx H_0$. Like in the flat potential case, there also exists an unstable de Sitter solution, now sitting at $H\approx \max(m, H_0)$.

{Before we conclude, a few comments are in order regarding the assumptions used in this work. In particular,} the present analysis is restricted to the local-potential approximation and to
homogeneous semiclassical gravity. It does not include metric
perturbations, graviton loops, stochastic noise or higher-derivative
operators. Accordingly, the stability discussed above should be
understood as stability within the homogeneous dynamical system derived
in this work. Since our aim is the deterministic evolution of the
homogeneous effective average action rather than the computation of
general real-time correlation functions, we do not formulate the problem
on a Schwinger--Keldysh contour. The equations obtained within the
present truncation are time-reversal invariant; genuinely dissipative
effects and more general nonequilibrium observables would instead require
an in-in treatment~\cite{Donath:2024utn}.

{Regarding the choice of cutoff function, the one used in the present work is the simplest one which leads to consistent Wilsonian flows.
Another choice would not change the qualitative results, but could slightly modify them quantitatively. Hence making precise phenomenological predictions would necessitate a study of the dependence on this cutoff function. We note though, that we have the freedom in the choice of initial potential, which could help compensate the change of cutoff function,
in order to lead to the same effective potential at the scale $H$. It would be interesting to look for a combination (initial potential/cutoff function) which leads to invariant physics, and this is left for future work. }

{Another important aspect of our work is that the particular choice of coarse-graining scale $\kappa=H$ is the cornerstone of our Wilsonian cosmology framework and corresponds to a choice of {\em synchronisation} between the RG clock and the gravitational one. 
Indeed, it is easy to see from the equations of motion we have derived, that depending on the choice of coarse-graining scale $\kappa$, the dynamics of $X$ can differ significantly. However, similarly to choosing different cutoff functions, this difference corresponds to the description of two intrinsically different systems: an initial choice of potential for a fixed value of $H$ but two different values of Wilsonian scales corresponds to the choice of two initially distinct theories.
}

Finally, we plan to develop studies related to the spinodal instability, 
which occurs with concave potentials $(\partial_\phi^2 U<0)$ and is a bit similar to a tachyonic instability.
These potentials are particularly relevant to inflationary cosmology, since they allow a long enough slow-roll period. 
It has been shown in \cite{Alexandre:1998ts} that this instability, in flat spacetime at least, is avoided with non-trivial saddle points in the path integral representing the Wilsonian blocking procedure, and the extension to cosmological background is a promising direction to follow \cite{Boyanovsky:1999wd}.

In most of the examples presented in the massive scalar case, it is remarkable that the Hubble scale undergoes damped oscillations when approaching the attractor, with excursions during which $H'>0$. In the simple cosmological setup studied throughout this article, this has a direct interpretation in terms of the
effective cosmological equation of state
\be
w_{\rm eff}
=
-1-\frac{2H'}{3H}
=
-1+\frac{2}{3}\epsilonH~,
\ee
which is still valid in our case, since it is independent of the details of the potential.
Consequently, periods with \(H'>0\) correspond to transient phantom-like
excursions with \(w_{\rm eff}<-1\)~\cite{Caldwell:1999ew}. Remarkably, this
behaviour arises here without introducing any new physics, but by requiring a self-consistent running of the effective potential with cosmological energy scales. This observation may be particularly relevant for late-time cosmology,
given that recent DESI DR2 analyses, when combined with CMB and supernova data, strengthen the indications for an evolving dark-energy sector and show a preference for histories involving a crossing of the phantom divide
\(w=-1\)~\cite{DESI:2025zgx,DESI:2025fii}. Assessing whether the present cosmological scenarios can accommodate recent cosmological data would also require the inclusion of matter and radiation, the reconstruction of \(w_{\rm DE}(z)\), 
and a perturbative analysis of spatial fluctuations establishing the absence of ghost and gradient instabilities~\cite{Vikman:2004dc}. 
These questions will be addressed in a dedicated study of dark energy within the Wilsonian Cosmology framework.

{More generally, by consistently including quantum effects in the discussion of scalar dynamics in cosmology, this Wilsonian Cosmology provides a framework that is suitable to study various scalar cosmological theories. In particular, cosmic inflation theories (already studied in a preliminary version of this formalism in Ref.~\cite{Alexandre:2025ixz}), axion dark matter theories, and quintessence theories are scenarios that will need to be scrutinised within this framework, which we will present in further studies.}

\section*{Acknowledgements}
The authors would like to thank Christian Byrnes, Laura Iacconi, Eugene Lim, Daniel Litim, and Manuel Reichert for helpful discussions.  The work of JA and LH was supported by the STFC under UKRI grant ST/X000753/1. The work of SP was funded by the Deutsche Forschungsgemeinschaft (DFG, German Research Foundation) under Germany’s Excellence Strategy – EXC 2094 – 390783311. LH acknowledges support by Institut Pascal and the P2I axis of the Graduate School of
Physics during the Paris-Saclay Astroparticle Symposium
2025, as well as the CNRS IRP UCMN. LH and JA would also like to thank the workshop {\it Cosmological Probes of the Early Universe}, King's College London, and the Global Engagement Partnership Fund.

\appendix
\numberwithin{equation}{section}

\section{Derivation of the flow equation}\label{derivation}

Starting from the action
\bea
&&S_k[\Phi]\\
&=&\int dt\sqrt{-g}\int d^3x\left(-\frac{g^{\mu\nu}}{2}\partial_\mu\Phi\partial_\nu\Phi-\frac{\xi}{2}R\Phi^2-U_i(\Phi)\right)\nn
&&-\frac{1}{2}\int dt \sqrt{-g}\int_p\varphi(t,\vec p)\varphi(t,-\vec p)~\Big(C_k(p^2)-i\varepsilon\Big)~,\nonumber
\eea
the partition function for the quantum field $\varphi$ on the background $\phi(t)$ is 
\bea \label{eq:eq00App}
&&Z_k[j]\equiv \exp\Big(iW_k[j]/\hbar\Big)\\
&=&\int{\cal D}[\varphi]\exp\left((i/\hbar)S_k[\Phi]+(i/\hbar)\int dt \sqrt{-g}\int d^3x ~j\varphi\right)\nonumber
\eea
The background fluctuation field with momentum $\vec p$ is
\bea
\varphi_b(t,\vec p)&=&\left<\varphi(t,\vec p)\right>\equiv\frac{1}{\sqrt{-g}}\frac{\delta W_k}{\delta j(t,-\vec p)}\\
&=&-\frac{i\hbar}{\sqrt{-g}Z_k}\frac{\delta Z_k}{\delta j(t,-\vec p)}~,\nonumber
\eea
and the second functional derivative of $W_k$ is 
\bea
&&\frac{\hbar~\delta^2 W_k}{\delta j(t,-\vec p)\delta j(t',-\vec q)}\\
&=&i\sqrt{g(t)g(t')}\Big(\left<\varphi(t,\vec p)\varphi(t',\vec q)\right>-\varphi_b(t,\vec p)\varphi_b(t',\vec q)\Big)~.\nonumber
\eea
The 1PI effective action is defined as the Legendre transform of $W[j]$, with the cutoff function term removed,
\bea
&&\Gamma_k[\phi+\varphi_b]=W_k[j]-\int dt' \sqrt{-g}\int d^3x ~j\varphi_b \\
&&+\frac{1}{2}\int dt' \sqrt{-g}\int_p \varphi_b(t',\vec p)\varphi_b(t',-\vec p)(C_k-i\varepsilon)~,\nonumber
\eea
and $j$ should be understood as a functional of $\varphi_b$. $\Gamma_k$ has the following functional derivatives
\bea
&&\frac{1}{\sqrt{-g}}\frac{\delta\Gamma_k}{\delta\varphi_b(t,p)}\\
&=&-j(t,-\vec p)+\varphi_b(t,-\vec p)(C_k-i\varepsilon)\nonumber
\eea
and
\bea
&&\frac{\delta^2\Gamma_k}{\delta\varphi_b(t,\vec p)\delta\varphi_b(t',\vec q)}\\
&=&\sqrt{-g(t)}(C_k-i\varepsilon)\delta(\vec p+\vec q)\delta(t-t')\nn
&&-\sqrt{g(t)g(t')}\left(\frac{\delta^2W_k}{\delta j(t,-\vec p)\delta j(t',-\vec q)}\right)^{-1}~.\nonumber
\eea
The evolution of the effective action with $k(t)$ is obtained from the functional derivative
\bea\label{dGammadkApp}
&&\frac{\delta\Gamma_k}{\delta k(t)}=\frac{\delta W}{\delta k(t)}
+\frac{\sqrt{-g}}{2}\int_p \varphi_b(t,\vec p)\varphi_b(t,-\vec p)\partial_k C_k\\
&=&\frac{\sqrt{-g}}{2}\int_p
\Big(\varphi_b(t,\vec p)\varphi_b(t,-\vec p)-\left<\varphi(t,\vec p)\varphi(t,-\vec p)\right>\Big)\partial_k C_k\nn
&=&i\frac{\sqrt{-g}}{2}\int_p\int_q\delta(\vec p+\vec q)\int dt' \delta(t-t')
~{\cal O}(t,t',\vec p,\vec q)~\partial_k C_k~,\nonumber
\eea
where 
\bea\label{OApp}
&&{\cal O}(t,t',\vec p,\vec q)\\
&=&\frac{\hbar}{\sqrt{g(t)g(t')}}\frac{\delta^2W_k}{\delta j(t,-\vec p)\delta j(t',-\vec q)}\nn
&=&\hbar\Bigg(\sqrt{-g(t)}(C_k-i\varepsilon)\delta(\vec p+\vec q)\delta(t-t')\nn
&&~~~~~~~~~~-\frac{\delta^2\Gamma_k}{\delta\varphi_b(t,\vec p)\delta\varphi_b(t',\vec q)}\Bigg)^{-1}~.\nonumber
\eea
In the LPA we have
\bea
&&\Gamma_k[\phi+\varphi_b]\\
&=&\frac{1}{2}\int dt\sqrt{-g}\int d^3x\Big(\big(\dot\phi(t)\big)^2+2\dot\phi(t)\dot\varphi_b(t,\vec x)\Big)\nn
&&+\frac{1}{2}\int dt\sqrt{-g}\int_p~\dot\varphi_b(t,\vec p)\dot\varphi_b(t,-\vec p)\nn
&&-\frac{1}{2}\int dt\sqrt{-g}\int_p~\frac{p^2}{a^2}\varphi_b(t,\vec p)\varphi_b(t,-\vec p)\nn
&&-\int dt\sqrt{-g}\int d^3x\left(\frac{\xi}{2}R(\phi+\varphi_b)^2 + U_k(\phi+\varphi_b)\right)~,\nonumber
\eea
For a vanishing source $j=0=\varphi_b$, the evolution of the effective action evaluated at the background field is then given by
\be\label{dGammadkbisApp}
\frac{1}{\sqrt{-g}}\frac{\delta\Gamma_k[\phi]}{\delta k(t)}=-V\partial_kU_k(\phi)~,
\ee
where $V$ is the space volume. The operator (\ref{OApp}) is therefore
\bea\label{OtpqApp}
&&{\cal O}(t,t',\vec p,\vec q)=\hbar\delta(t-t')\delta(\vec p+\vec q)\Bigg(\partial_t(\sqrt{-g}\partial_t)\\
&&~~~+\sqrt{-g}\left[p^2a^{-2}+\xi R+\partial^2_\phi U_k(\phi)+C_k-i\varepsilon\right]\Bigg)^{-1}~.\nonumber
\eea 
For the previous step we used the fact that the distributions $\delta(t-t')$ and $\delta(\vec p+\vec q)$ are identical to their inverse.
The derivatives $\partial_t(\sqrt{-g}~\partial_t)$ acting on the time-dependent quantities, when taking the inverse, 
introduces poles in the complex $p$-plane. To avoid this, we perform the Wick rotation $t\to it$ 
and, in the limit $\varepsilon\to0$, the resulting Euclidean version of the operator (\ref{OtpqApp}) acting on the cutoff function is
\bea
&&{\cal O}_E(t,t',\vec p,\vec q)\partial_k C_k\\
&=&\hbar\delta(t-t')\delta(\vec p+\vec q){\cal D}_E^{-1}\left(a^{-3}\partial_k C_k\right)~,\nonumber
\eea
where the Euclidean differential operator is
\be
{\cal D}_E=-a^{-3}\frac{d}{dt}\left(a^3\frac{d}{dt}\right)+p^2a^{-2}-\xi R+\partial^2_\phi U_k(\phi)+C_k~.\nonumber
\ee
After the Wick rotation, Eq.~(\ref{dGammadkbisApp}) should involve the Euclidean effective action
\be
\frac{1}{\sqrt{-g}}\frac{\delta\Gamma_k[\phi]}{\delta k(t)}\to 
\frac{i}{\sqrt{g_E}}\frac{\delta\Gamma_k^E[\phi]}{\delta k(t)}=iV\partial_kU_k(\phi)~,
\ee
such that the evolution equation for the effective action is finally
\be\label{dkUApp}
\partial_k U_k(\phi)=\hbar\frac{T^{-1}}{2}\int_p{\cal D}_E^{-1}\Big(a^{-3}\partial_kC_k\Big)~,
\ee
where 
\be
T^{-1}\equiv \lim_{t'\to t}\delta(t-t')~~~~\mbox{and}~~~~
V=\lim_{\vec p+\vec q\to0}\delta(\vec p+\vec q)~.
\ee
With the Litim cutoff for the physical scales $p/a$ and $\kappa\equiv k/a$, we have
\be
C_k(p^2)=a^{-2}(k^2-p^2)\Theta(k^2-p^2)~,
\ee
in which case the evolution equation (\ref{dkUApp}) becomes 
\bea\label{flowEApp}
\partial_\kappa U_\kappa(\phi)&=&\hbar\frac{T^{-1}}{2}{\cal D}_E^{-1}\int_p a^{-3}\partial_kC_k\\
&=&\hbar\frac{aT^{-1}}{6\pi^2}{\cal D}_E^{-1}\left(a^{-1}\kappa^4\right)~,\nonumber
\eea
where 
\be\label{DEApp}
{\cal D}_E=-\frac{d^2}{dt^2}-3H\frac{d}{dt}+\kappa^2-\xi R+ \partial^2_\phi U_\kappa(\phi)~,
\ee
and $\kappa$ is defined here with imaginary time. We then multiply by the inverse operator ${\cal D}_E$ on both sides of the equation, to obtain
\be\label{stepApp}
a{\cal D}_E\left(a^{-1}T\partial_\kappa U_\kappa(\phi)\right)=\frac{\hbar\kappa^4}{6\pi^2}~,
\ee
such that we turn a potential series in time derivatives into a differential equation for $\partial_\kappa U_\kappa(\phi)$.
Going back to real time, we obtain then
\be\label{flow0App}
{\cal D}\left(T\partial_\kappa U_\kappa(\phi)\right)=\frac{\hbar\kappa^4}{6\pi^2}~,
\ee
where we define
\bea\label{defDApp}
{\cal D}&\equiv&\frac{d^2}{dt^2}+H\frac{d}{dt}+\kappa^2+{\cal M}^2\\
{\cal M}^2&\equiv&\partial^2_\phi U_\kappa(\phi)+(\xi-1/6)R~,\nonumber
\eea
and $\kappa$ is here defined with real time.

\section{Review of the derivative expansion in curved spacetime}\label{review}

A common strategy to incorporate quantum corrections in slow-roll inflation is to work with the (one-loop) 1PI effective action in a slowly varying background. In practice, this is implemented via a local (derivative) expansion of the one-loop effective action in curved spacetime, often organised through heat-kernel techniques. In this section we review the curvature-resummed heat-kernel derivation of the adiabatic one-loop effective potential \cite{Parker:1984dj,Jack:1985mw}, following
standard treatments in curved spacetime. 

We start in Euclidean signature and consider the bare matter action
\be\label{eq:SE_adiabatic}
S_E[\Phi,g]=\int d^d x\,\sqrt{g}\left[
\frac{1}{2}g^{\mu\nu}\partial_\mu\Phi\,\partial_\nu\Phi
+\frac{\xi}{2}R\,\Phi^2
+U_i(\Phi)
\right]~,
\ee
where $U_i$ is the initial (bare) potential. We expand around a background configuration $\Phi=\phi+\delta\phi$, and keep the quadratic action
for $\delta\phi$,
\be
S_E[\phi+\delta\phi,g]=S_E[\phi,g]+\frac{1}{2}\int d^d x\,\sqrt{g}\;\delta\phi\;{\cal D}_E\;\delta\phi+\cdots~,
\ee
where the fluctuation operator reads,
\be\label{eq:DE_def}
{\cal D}_E\equiv -\Box_E + Q(\phi,g)~,
\qquad
Q(\phi,g)\equiv U_i''(\phi)+\xi R~,
\ee
and $\Box_E=g^{\mu\nu}\nabla_\mu\nabla_\nu$.
At one loop, the 1PI effective action reads (see, for example Refs. \cite{Parker:2009uva,Hu:2020luk})
\bea\label{eq:Gamma1_def}
\Gamma_E[\phi,g]&=&S_E[\phi,g]+\Gamma_E^{(1)}[\phi,g]+{\cal O}(\hbar^2),\\
\Gamma_E^{(1)}[\phi,g]&=&\frac{\hbar}{2}\Tr\ln{\cal D}_E~.
\eea
For general background configurations, the one loop effective action is unknown. However, an
approximated expression in the case of slowly varying background fields  can be computed using the proper-time formalism as follows (see Refs. \cite{Hu:1984js,Markkanen:2018bfx} for a detailed explanation).

Using the Schwinger proper-time representation of the functional determinant, one writes 
\be\label{eq:proper_time}
\Tr\ln{\cal D}_E = -\int_0^\infty\frac{ds}{s}\,\Tr\!\left(e^{-s{\cal D}_E}\right)~,
\ee
so that
\bea\label{eq:Gamma_proper_time}
\Gamma_E^{(1)}[\phi,g]
&=& -\frac{\hbar}{2}\int_0^\infty\frac{ds}{s}\;K(s),\\
K(s)&\equiv& \Tr\!\left(e^{-s{\cal D}_E}\right)
=\int d^d x\,\sqrt{g}\;K(s;x,x)~,
\eea
where $K(s;x,x')\equiv\langle x|e^{-s{\cal D}_E}|x'\rangle$ is the heat kernel \cite{Vassilevich:2003xt}. In the adiabatic regime, one uses the small-$s$ expansion of $K(s;x,x)$.
A more convenient expansion is the heat kernel expansion in its  curvature-resummed form 
\be\label{eq:Rsummed_HK}
K(s;x,x)=\frac{e^{-{\cal M}^2 s}}{(4\pi s)^{d/2}}\sum_{k=0}^{\infty}b_k(x)\,s^k~,
\ee
with
\be\label{eq:M2_def}
{\cal M}^2\equiv Q-\frac{1}{6}R=\partial_\phi^2U_i(\phi)+\Big(\xi-\frac{1}{6}\Big)R~.
\ee
Note that this is the same  combination that will appear in our coarse-grained framework. The first coefficients read \cite{Parker:2009uva}
\be
b_0=1,\qquad b_1=0,
\ee
and the next term collects curvature and derivative invariants 
\bea\label{eq:b2_def}
b_2 &=&\frac{1}{180}R_{\mu\nu\rho\sigma}R^{\mu\nu\rho\sigma}-\frac{1}{180}R_{\mu\nu}R^{\mu\nu}\\
&&+\frac{1}{30}\Box_E R-\frac{1}{6}\Box_E Q~.\nonumber
\eea
Truncating \eqref{eq:Rsummed_HK} at a finite order $N$ defines the ($R$-summed) adiabatic derivative expansion. Inserting \eqref{eq:Rsummed_HK} into \eqref{eq:Gamma_proper_time} and truncating at $k\le N$ gives
\be \begin{aligned}
\label{eq:Gamma_trunc}
\Gamma_E^{(1)}[\phi,g]
=&
-\frac{\hbar}{2}\int d^d x\,\sqrt{g}\,(4\pi)^{-d/2}\times\\
&\sum_{k=0}^{N} b_k(x)\int_0^\infty ds\; s^{k-1-d/2}\,e^{-{\cal M}^2 s}~.
\end{aligned}\ee
The proper-time integral is straightforward:
\be\label{eq:s_integral}
\int_0^\infty ds\; s^{k-1-d/2}\,e^{-{\cal M}^2 s}
=({\cal M}^2)^{d/2-k}\,\Gamma\!\left(k-\frac{d}{2}\right)~,
\ee
hence
\be \begin{aligned}
\label{eq:Gamma_dim}
\Gamma_E^{(1)}[\phi,g]
=&
-\frac{\hbar}{2}\int d^d x\,\sqrt{g}\,(4\pi)^{-d/2}\times\\
&\sum_{k=0}^{N} b_k(x)\,({\cal M}^2)^{d/2-k}\,
\Gamma\!\left(k-\frac{d}{2}\right)~.
\end{aligned}\ee
The UV divergences arise from the poles of $\Gamma(k-d/2)$ as $d\to 4$.
In dimensional regularisation, $d=4-\epsilon$, and after renormalisation in the $\overline{\rm MS}$ scheme,
 yields the standard local result 
\be \begin{aligned}\label{eq:Gamma_ren_local}
\Gamma_{E,\rm ren}^{(1)}[\phi,g]
=&
\hbar\int d^4 x\,\sqrt{g}\Bigg[
\frac{{\cal M}^4}{64\pi^2}\left(\ln\frac{{\cal M}^2}{\mu^2}-\frac{3}{2}\right)\\
&+\frac{b_2}{32\pi^2}\ln\frac{{\cal M}^2}{\mu^2}
+\cdots
\Bigg]~,
\end{aligned}\ee
where $\mu$ is the renormalisation scale and the ellipsis stands for higher-derivative terms.
The heat-kernel computation leading to Eq.~\eqref{eq:Gamma_ren_local} is performed in Euclidean signature. For the terms retained in the adiabatic
expansion, one may analytically continue back to Lorentzian signature and evaluate the resulting invariants
on a Lorentzian FLRW background. 

In the Local Potential Approximation, the adiabatic one-loop effective potential is defined as the local
potential energy-density term appearing in the effective action (i.e. $\Gamma^{(1)}_{\rm ren}[\phi,g]\equiv-\int d^4x\sqrt{-g}\,U^{(1)}_{ren}$ once derivative operators are neglected).
Keeping only the $b_0$ contribution in the $R$-summed expansion then gives
\be\label{eq:Uad_final}
U^{(1)}_{ren}=U_{i}
+\hbar\,\frac{{\cal M}^4}{64\pi^2}\left(\ln\frac{{\cal M}^2}{\mu^2}-\frac{3}{2}\right)~,
\ee

In  subsection  \ref{subsec:recovering1PI} we showed that our coarse-grained flow equation in FLRW  reproduces the result above 
in the strict adiabatic limit.

\bibliographystyle{apsrev4-1}
\bibliography{main}

\end{document}